\documentclass[12pt,preprint]{aastex}

\begin{document}

\title{L-dwarf variability: $I$-band observations}
\author{Christopher R. Gelino, Mark S. Marley\altaffilmark{1}, Jon A.
Holtzman}
\affil{Department of Astronomy, New Mexico State University, Las Cruces,
NM, 88003}
\email{crom@nmsu.edu, mmarley@mail.arc.nasa.gov, holtz@nmsu.edu}

\author{Andrew S. Ackerman}
\affil{NASA Ames Research Center, Moffett Field, CA, 94035}
\email{ack@sky.arc.nasa.gov}

\and

\author{Katharina Lodders}
\affil{Planetary Chemistry Laboratory, Department of Earth and Planetary
Sciences, Washington University, St. Louis, MO, 63130}
\email{lodders@levee.wustl.edu}

\altaffiltext{1}{NASA Ames Research Center, Moffett Field, CA, 94035}


\begin{abstract}
We report on the results of an $I$-band photometric variability survey of
eighteen L dwarfs.  We find that seven exhibit 
statistically significant variations above the 95.4\% confidence level with
root-mean-square scatter (including photometric errors) between 0.010 and 0.083
mag.  Another five targets have variability probabilities $\approx$80\%,
suggesting that these are likely variable objects. 
Three of the variable objects display significant peaks in a CLEAN 
periodogram that are several times higher than the noise.
The period for 2MASS 0345+25 is clearly
not intrinsic to the object and can be dismissed.  The periods 
found for 2MASS 0746+20AB
and 2MASS 1300+19 are unique but longer than those periods
likely from rotation velocity measurements and they do not 
represent periodic behavior in the light curve that persists
through the entire data set.  These observations suggest
that we are not observing the rotation modulation of a long-lived albedo 
feature.  Instead, rapid evolution of atmospheric features
is likely causing the non-periodic variability. The remaining variable objects 
show no prominent features in their light curves, suggesting even more rapid 
evolution of atmospheric features.   We argue against the
existence of magnetic spots in these atmospheres and favor the idea that
non-uniform condensate coverage is responsible for these
variations.  The magnetic Reynolds number in the atmosphere of L dwarfs is 
too small to support the formation of magnetic spots.  In contrast, silicate 
and iron clouds are expected to form in the photospheres
of L dwarfs. Inhomogeneities in such cloud decks and the evolution of the
inhomogeneities can plausibly produce the observed photometric variations.  
\end{abstract}

\section{Introduction}
Large numbers of L dwarfs \citep{kir99,mart99,bas00,kir00} are being 
discovered by surveys
such as the 2 Micron All Sky Survey (2MASS) and the Sloan Digital Sky Survey 
(SDSS).  Despite their growing number, less than 10\% of known L dwarfs have
been observed to determine whether or not they 
are photometrically variable \citep*{tin99,bai99,bai01a,cla02}.
Photometric monitoring of L dwarfs, and brown
dwarfs in general, not only explores why some objects are variable while
others are not, but it can also help constrain rotation periods or 
timescales for the evolution of surface features.  Multicolor photometric and 
spectroscopic monitoring could 
even be used to identify the nature of those features.

\citet{tin99} presented the first attempt to detect clouds in brown
dwarf atmospheres.  They observed the M9 brown dwarf LP944$-$20 and the L5
brown dwarf DENIS 1228$-$15 through two narrow-band filters chosen
to detect changes in TiO absorption.  The changes in the TiO band
strength were presumed to indicate changes in the opacity, which occurs
when TiO is depleted through condensation.  They found that
LP944$-$20 was variable, but DENIS 1228$-$15 was not.  The
authors speculated that the passage of clouds over the disk of LP944$-$20
produced small changes in the brightness temperature in their narrow-band 
filters and caused small variations ($\sim$0.04 magnitude).  However, the 
lack of variability in their L dwarf does not exclude the possibility of 
clouds in that object since the variations observed in LP944$-$20 had smaller 
amplitudes than the errors for DENIS 1228$-$15.

\citet{bai99} conducted a variability search 
in the broad-band $I$ filter and found evidence of variability in
the L1.5 dwarf 2MASS 1145+23.  The object displayed $\sim$0.04
magnitude variations that repeated with a period of 7.1 hours.
In an expanded study of 21 L and M dwarfs
\citet{bai01a,bai01b} found that
over half of their sample exhibited statistically significant variations
with root-mean-square (RMS) scatter between 0.010 and 0.055 magnitude and time 
scales of 0.4 to 100
hours.  They were unable to find periodic light curves for many of the
variables. 2MASS 1145+23, however, exhibited
variability with a period of 11.2~hours.  They suggested that
evolving surface features, possibly dust clouds or
magnetic spots, were responsible for the change in period. 
A similar varying period has also been observed in the M9.5 dwarf
star BRI0021 by \citet*{mart01}.

Photometric variability has recently been reported for the L2 dwarf Kelu-1
\citep{cla02}.  Like \citet{tin99}, these authors use a special filter 
positioned at a region sensitive to changes in TiO and CrH absorption.  Kelu-1
displayed small ($\sim 1.1\%$) peak-to-peak variations that phased well to a 
period of 1.8 hours.

Either magnetic spots or clouds could plausibly
be associated with the observed variability.  Magnetic fields have
been measured for several M dwarfs and estimates of their strength
are a few kG \citep{saar94,joh96}, with surface filling factors 
generally $>50\%$.  Periodic photometric
variations have been observed in only
a few M dwarfs, suggesting that either the surfaces of these objects are
completely covered with spots or that the spots are few in number and
uniformly distributed \citep*{haw00}.  In addition, \citet{bon95}
reports that no good data exist to support the scenario of cyclic,
organized spots in M dwarfs cooler than M0.  

For the L dwarfs, clouds are a reasonable potential source
of variability.  Iron, enstatite, and forsterite are
the most abundant species expected to condense at the atmospheric
temperatures and pressures characteristic of the L dwarfs 
\citep{lod99,bur99}.
Once condensed, the species likely settle into discrete, optically
thick, cloud decks, with optically thicker clouds arising in progressively
later L dwarfs \citep{marl00,ack01}.  Since
the atmospheric circulation pattern of most
L dwarfs is likely similar to that of Jupiter
\citep{sch00}, it is not unreasonable to expect that many
L dwarfs also have a banded appearance.  Any large  inhomogeneities
(thicker clouds or clearings in the cloud deck) could then produce
a photometric signal.  \citet{gel00} show that if Jupiter were to be
 observed in thermal emission as an
unresolved point source, the Great Red Spot would provide a
photometrically detectable signal.  If L dwarfs have similar cloud
features to Jupiter, then it is plausible that they may also
exhibit photometric variations.  

We are conducting a photometric monitoring program of L dwarfs in the $I_C$
band.  In this study we present
the light curves for several L dwarfs showing statistically significant 
variability.  We discuss which
of these display significant periodicity and why the others do not.
We provide theoretical calculations that reject the
magnetic spot hypothesis for the origin of the variation and argue in favor of 
clouds.

\section{Observations}

\subsection{Target Selection}
Our target list was derived from a compilation of all published
spectroscopically determined L dwarfs.  We 
searched the list for all L dwarfs above $\delta=-10\degr$ and brighter 
than $I\approx18$.  For the objects whose $I$ magnitudes had not been measured 
we estimated their brightness in that filter based upon the objects' spectral 
types and the $I-J$ colors from \citet{kir99}.  Better estimates of these $I$ 
magnitude were made after observations using instrumental magnitudes
and $I-J$ colors.

At the start of this project there were 24 objects that fit these criteria.  
Table~\ref{tab-list} lists the 18 L dwarfs in our sample that were observed.
Except for an L4, we have at least one object for each decimal subclass between
L0 and L4.5, inclusive.
All are discoveries from the 2MASS survey \citep{giz00,kir99,kir00} and are 
field objects.  Their $I$-band magnitudes range from 15.11 mag (measured) 
to $\sim$18.1 mag (estimated).
Finally, exactly half of the sample has measurable H$\alpha$ emission.

2MASS 0746+20AB and 2MASS 1146+22AB are close visual binaries as seen with 
Keck and the Hubble Space Telescope \citep{koe99,reid01} and are
unresolved in our observations.  \citet{reid99} state that Keck echelle
spectroscopy indicate 2MASS 0345+25 is a spectroscopic binary.
However, additional spectra of 2MASS 0345+25 can neither 
confirm nor reject the possibility
of a binary system (Reid 2001, private communication).  Only two other targets
were observed for binarity (2MASS 0036+18 and 2MASS 1338+41; Reid et al. 2001),
but none was found.  However, nearly 25\% of all L dwarfs observed in binary
studies are shown to be binaries.  Therefore, the probability that other 
binaries exist in our sample is quite high.

\subsection{Data Acquisition}
We used the New Mexico State University 1m telescope at Apache Point 
Observatory for our observations. 
The telescope was equipped with an Apogee 512$\times$512 pixel, 
thermo-electrically 
cooled CCD with a pixel scale of 0.8 arcsec/pixel and was operated 
robotically.  For each cycle the telescope was focused 
on a bright star and the script chose one to 
three targets that had airmasses between 1.03 and 1.9.
After slewing to a target, the telescope pointing was 
refined by slewing to a nearby bright standard star and centering the star
in the CCD.  The telescope reslewed to the target and found a guide star from
the USNO-SA2.0 catalog \citep{mon98}.  On most nights three consecutive 5 
minute exposures 
were taken in the Cousins $I$ band for each target; no other filters or
exposure times were used.  This cycle was repeated as many times as possible
throughout the night.  Each target was usually observed every 1-2 hours in a
given night, depending on the target's hour angle and priority.

Observations for this program started on 2000 October 30 and ended on 
2001 June 20.  During this 
period the telescope was used as much as possible for this program.
Observations were taken at all moon phases and under a variety of seeing
conditions.
Over 3700 science images (not including positional calibration frames and 
focus runs) 
were obtained, resulting in over 300 hours of on-sky time for science 
observations.  As with any observing program, 
time was occasionally lost due to inclement weather and engineering 
problems.  The longest off-sky period was approximately 1 month from 
mid-December to mid-January.

\subsection{Data Reduction}
The images were reduced using procedures in the XVISTA astronomical image
reduction and analysis package.  All images were first dark subtracted and 
then 
flat fielded.  Finally, a correction was applied to reduce the effect of night
sky emission line fringing in the frames.  These steps are discussed in more 
detail below.

\subsubsection{Dark Subtraction}
Dark current is an additive noise that depends strongly on the operating 
temperature of the CCD and the exposure time.  The amount of charge accumulated
increases with the exposure time; the rate of charge production increases with
increasing temperature.  We limited our thermo-electrically cooled CCD to be 
set at only three possible temperatures ($-$40, $-$45, and $-$50$^\circ$~C), 
dependent upon the ambient temperature.  The dark level was typically $<1\%$ 
of the total background for observations taken at $-$50$^\circ$~C and $<3\%$ 
for observations at $-$40$^\circ$~C. Consequently, dark current was an 
important source of background in our images.

We constructed a master dark from thirty 5-minute dark exposures obtained in 
late November taken at a CCD temperature of $-$50$^\circ$~C. (The resulting
error in photometry caused by using this dark frame with science frames 
obtained at the other temperatures was well below 1\%.)  The frames were 
combined using a median average and divided by the exposure time to yield a 
dark frame in counts per second.  The dark frame multiplied by the exposure 
time was subtracted from each image.

\subsubsection{Flat Field}
The division of the flat field removes the pixel-to-pixel variations of the
quantum efficiency.  As with the dark frame, we used the same flat field for
all images.  This flat was constructed from about 12 individual twilight
flat field frames taken around the same dates as the dark frames.  No major 
changes to the telescope or the instrument configuration were made 
during the entire observing program.  Nonetheless, we tested the flat obtained
in November with a set of flats obtained at the end of the program.  There was
an average deviation between the two flats of 0.5\% across the entire chip and 
a maximum deviation of about 3\% close to the edges.  Inspection of 
flat-fielded images did not show any prominent residual features
attributable to an improper flat.  However, to account for small flat field 
errors, we added an error of 0.005 mag in quadrature with the photometric 
errors.

\subsubsection{Fringe Subtraction}
Night sky emission lines can reflect several times inside a thinned CCD and
cause interference.  This interference is a source of coherent background in 
our science 
images and manifests itself as a fringe pattern present in the raw images. The 
pattern is stable but the amplitude of the fringes is highly variable
and depend on moonlight, the exposure time, cloud cover, and position in the 
sky.  Fringes are also an additive effect, unlike the multiplicative effect of
the flat field. Since many of our targets are faint compared to the background,
it is important that we correctly determine the pattern of the fringes and 
accurately compute the pattern's amplitude in the raw images.

Incidentally, there is no evidence of fringing in our twilight flats, 
presumably because the source of the light in the flats is scattered sun light.
The flux from this scattered 
light dominates over the night sky lines responsible for the fringing in the 
science frames.  

We assumed that the background of the images 
consisted of a normalized fringe frame representing the
interference component multiplied by some image-dependent factor $l_1$, a
constant continuum sky $l_0$, and brightness gradients along the X and Y axes
scaled by $l_2$ and $l_3$, respectively, to account for any moonlight or 
scattered light.  After a frame was flat fielded and 
dark subtracted a mask was placed on the frame to remove all pixels 5$\sigma$
above the sky.  The linear X and Y gradients were constructed and best fits
to the factors $l_0$, $l_1$, $l_2$, and $l_3$ were found for a 300$\times$300 
pixel box from the central region of the image.  The normalized fringe frame 
was multiplied by $l_1$ and subtracted from the image.  

We constructed the normalized fringe frame from a large set ($>50$) of
science frames that were either from different target fields or from fields in
which the stars had been significantly offset.  For a first guess at the 
fringe frame, we reduced the
individual frames with the flat field and dark frame.  Each frame was then
divided by its sky level, creating a set of normalized frames.  These frames
were median averaged to form a fringe frame.  The next iteration reduced the 
images with the dark, flat field, and
the first guess at the fringe frame so that the four components of the 
background, most importantly $l_0$ and $l_1$,
could be found. Next, we reduced the raw frames again, this time using only 
the flat and dark.  These frames were normalized by subtracting the sky 
continuum $l_0$ and dividing by the fringe level $l_1$.  They were then
median averaged to create a new fringe frame.  This process was iterated 
three more times, with each iteration using the fringe frame from the
previous iteration.  The final result was a fringe frame normalized to
a mean value of 0.

This method for the fringe removal worked reasonably well, in the sense that in
most frames the fringe patterns were completely removed.  For the frames in 
which the fringes were not completely removed, the residual fringe level was 
significantly $<1\%$ the level of the background.  We added an error term, 
discussed below, to account for sky subtraction errors from an imperfect
fringe subtraction.

\subsection{Photometry}
We used aperture photometry to obtain the instrumental magnitudes of the
targets and references.  A circular aperture with a radius of 4 pixels was 
used to compute source fluxes; sky levels were determined from an annulus
between 10 and 24 pixels from the source center.  We performed differential
photometry with the reference being 
the average brightness of non-variable stars in the field.  Using a reference
that is the average of many stars increased the 
signal-to-noise ratio in our target light curves.  Furthermore, any color 
dependence on telluric extinction due to the difference in color between the
targets and references was minimized by obtaining our science 
frames at low airmass (nearly 80\% of our science frames were obtained at an 
airmass below 1.4 and 50\% were below 1.24 airmasses).

During the reduction process some frames were thrown out because there was 
an obvious problem with those frames.  The most common problems were streaked 
stars due to tracking or guiding errors and missing stars from cloudy skies 
or an automatic dome closure.  Frames with less 
obvious problems (e.g. being slightly out of focus) were allowed to go through 
the reduction process, but some were removed later as discussed below.

The first step in computing the mean reference is the determination of which
reference stars are non-variable.  Our ability to detect the variability of
an object depends
strongly on that object's photometric errors.  These errors are computed 
from the Poisonian error of the object, the sky noise error, and the error of
the mean sky level, which is taken to be 0.15\% of the mean sky value.
We use this last term to account for errors in the fringe subtraction. We 
find that its use increases the photometric errors in all objects.  The 
increase is largest for the
fainter objects, allowing for a more conservative estimate of variability in 
these objects.  We also add an error of 0.005 mag in 
quadrature with the photometric errors to account for changes in the flat field
that occurred throughout the course of the observations.  This additional
error is a considerable fraction of the total error in the brightest objects
and negligible in the faint ones.  Consequently, its effect on the 
detection of variability is only important for the bright objects.

After all the data are collected for a target, the data set is put through 
several filters to remove bad points.  
The first processing filter removes all frames in which the average 
FWHM (full width at half maximum) of the stellar profiles is $>3\arcsec$.
This ensures that we do not use frames where the stars are slightly elongated,
out of focus, or subject to poor seeing conditions.  In general, these high 
FWHM frames have larger photometric errors. 

The data for each reference star, $k$, is analyzed in a method similar 
to \citet{bai99}
to determine if the reference is stable and acceptable.  This is accomplished
by first computing the average flux for all references in frame $j$, 
excluding reference $k$, and converting it to a magnitude,

\begin{equation}
\overline{f_{kj}}=\frac{1}{n-1}\sum^{n}_{i=1,i\neq k} 10^{-0.4 m_{ij}}
\end{equation}
\begin{equation}
\overline{m_{kj}}=-2.5 \log(\overline{f_{kj}}),
\end{equation}

\noindent where $m_{ij}$ is the instrumental magnitude for reference $i$ 
in frame $j$, $n$ is the number of references. 
The difference in magnitude between reference $k$ and the mean 
reference is formed and $\chi^2$ is computed in order to determine the
probability that $k$ is variable,

\begin{eqnarray}
\Delta m_{kj}=m_{kj} - \overline{m_{kj}} \\
\chi_k^2=\sum_{j=1}^{N} \biggl(\frac{\Delta m_{kj} - \overline{\Delta m_k}}{\sigma_{kj}}\biggr)^2,
\end{eqnarray}
where
\begin{equation}
\overline{\Delta m_k} = \frac{1}{N}\sum_{j=1}^{N} \Delta m_{kj},
\end{equation}
$N$ is the number of frames, and $\sigma_{kj}$ is the photometric error 
associated with $\Delta m_{kj}$.

The $\chi^2$ statistic is computed for each reference and the probability
of variability ($p$) for the reference with the largest $\chi^2$ is calculated.
If $p>$95.4\% (a 2$\sigma$ detection), that reference is flagged as variable 
and removed from the list of references.  This process of computing the mean
reference and removing variable references is iterated until no more variable
references are found.  Finally, we calculate a mean reference for each
frame based solely on the non-variable references.

Using all frames we compute the median brightness of the mean 
reference.  We then remove any frame for which the brightness of the mean
reference deviates by more than 0.5 mag from the median brightness.  This
effectively removes frames that are heavily obscured by terrestrial clouds.

Using this set of good frames, we re-analyze each reference (even the ones 
previously flagged as variable) and check for variability.  As before the
non-variable references are combined to make a mean reference for a given
frame.  In most cases the set of good references were the same before and 
after the removal of faint frames.  Table~\ref{tab-sv}
presents how many good references were found for each L-dwarf field.

\section{Target Data Analysis}
\subsection{Statistical Analysis}
Differential magnitudes are computed for the target, using the mean reference 
as the reference ``star.''  The mean brightness of this differential light
curve is found and a $\chi^2$ test is used to determine how much the light
curve deviates from a constant value centered on the mean.  The value of 
$\chi^2$ and the number of degrees of freedom in the computation of 
$\chi^2$ are used to calculate
the probability that $\chi^2$ is obtained by chance.  If this probability
is $<0.046$ (i.e. $p>95.4$\%, a 2$\sigma$ detection),
then the target is flagged as variable.  Targets suspected of
being variable are searched for periodic variations using the method 
described below. 

It is important to note $\chi^2$ for an object and, consequently, the 
variability probability, depend strongly on the estimate of that object's 
errors.  These errors are difficult to accurately determine and any changes 
to our estimate of the errors results in a change in $\chi^2$.  Therefore, 
these results should be regarded as more of a 
gauge for variability than an absolute classification. 

\subsection{CLEAN Periodogram}
A CLEAN periodogram routine \citep{rob87} is used to search for periods in the 
data of the suspected variable objects.  This routine computes a ``dirty''
power spectrum, whose components consist of the spectrum of the frequency
components convolved with the observing window function derived from the 
temporal sampling of the data.  The power spectrum is ``CLEANed'' by 
subtracting the dominant peak convolved with the window function from the
dirty spectrum to produce a residual spectrum.  To aide in the stabilization 
of the routine, only a fraction, called the gain,
of the response from the dominant period is removed.  Next, the second highest
peak (multiplied by the gain) is removed from the residual spectrum.   
Subsequent peaks are removed in the same fashion until all CLEAN components 
are obtained.  This process is repeated for additional CLEANings.  
After a user--specified number of CLEANings are performed, the final 
spectrum is constructed based on the CLEAN components and the final residual 
spectrum.  For consistency in our period computations we use a gain of 0.5 and 
100 CLEANs for all variable targets.

The CLEAN routine removes most of the
``problem'' frequencies associated with the window function that are 
present in a Lomb-Scargle routine \citep{press92}.
The result is a routine that is 
better able to find significant periods present in the data.  The downside
to this method is that there is no indicator for the noise in the power 
spectrum.

\citet{bai01a} provide an algorithm for estimating the noise in a CLEAN
power spectrum.  This noise depends on the total amount of time in
large gaps (we define a large gap as any time between points $>$12 hours)
for the observing run.  The amount of time attributed to large gaps was 
usually considerable for our runs and resulted in very small power spectrum 
noise values.  Consequently, numerous peaks could occur in a given power
spectrum that are several hundred times higher than the computed noise.  This
makes it difficult to establish which peaks are significant and which are
meaningless.

To overcome this shortcoming, we attempt to estimate the noise by calculating
the power expected for a synthetic data set with the same temporal sampling 
and RMS scatter as the real data set.  The synthetic data set is created by 
randomly shuffling the magnitudes
in the real data set to other times in the data set.  The values of the 
magnitudes do not change, only the time at which they occur.  Any periodicity 
in the original data set is
erased when the data are randomized.  Therefore, the results of the CLEAN 
analysis for the synthetic data provide what powers can be expected for a set
a random data with a certain RMS scatter.  We take the mean value of the 
primary peak power from 1000 random shuffles as the noise level in the original
data set.  We calculate the primary peak power to noise ratio (PNR); data 
with periodic variations should have a PNR value several times greater than 1.

\section{Results}
\label{results}
The variability results for our sample are presented in Table~\ref{tab-sv}.
Included are the total number of nights each object was observed and the 
number of nights in the final data set, the number of good images used in the 
analysis and the total taken, the number of days the data sets
cover (t$_{max}$), the number of good references used for the mean reference
and the total number initially considered, the probility the L dwarf is 
variable ($p$), the 
standard deviation of the points from the mean level ($\sigma_{\rm RMS}$), and
the average error ($\overline{\sigma_m}$).

We have already defined the variable L dwarfs as those with
probabilities above 95.4\%.  This group constitutes 7 out of the sample of 18
targets.  The five objects 
with $p\sim$80\% are considered possibly variable, since they are likely to be 
viewed as variable with higher precision photometry.  The six objects with
$p<35\%$ clearly show no variability above the detection limits of this study.

\begin{figure}
\plotone{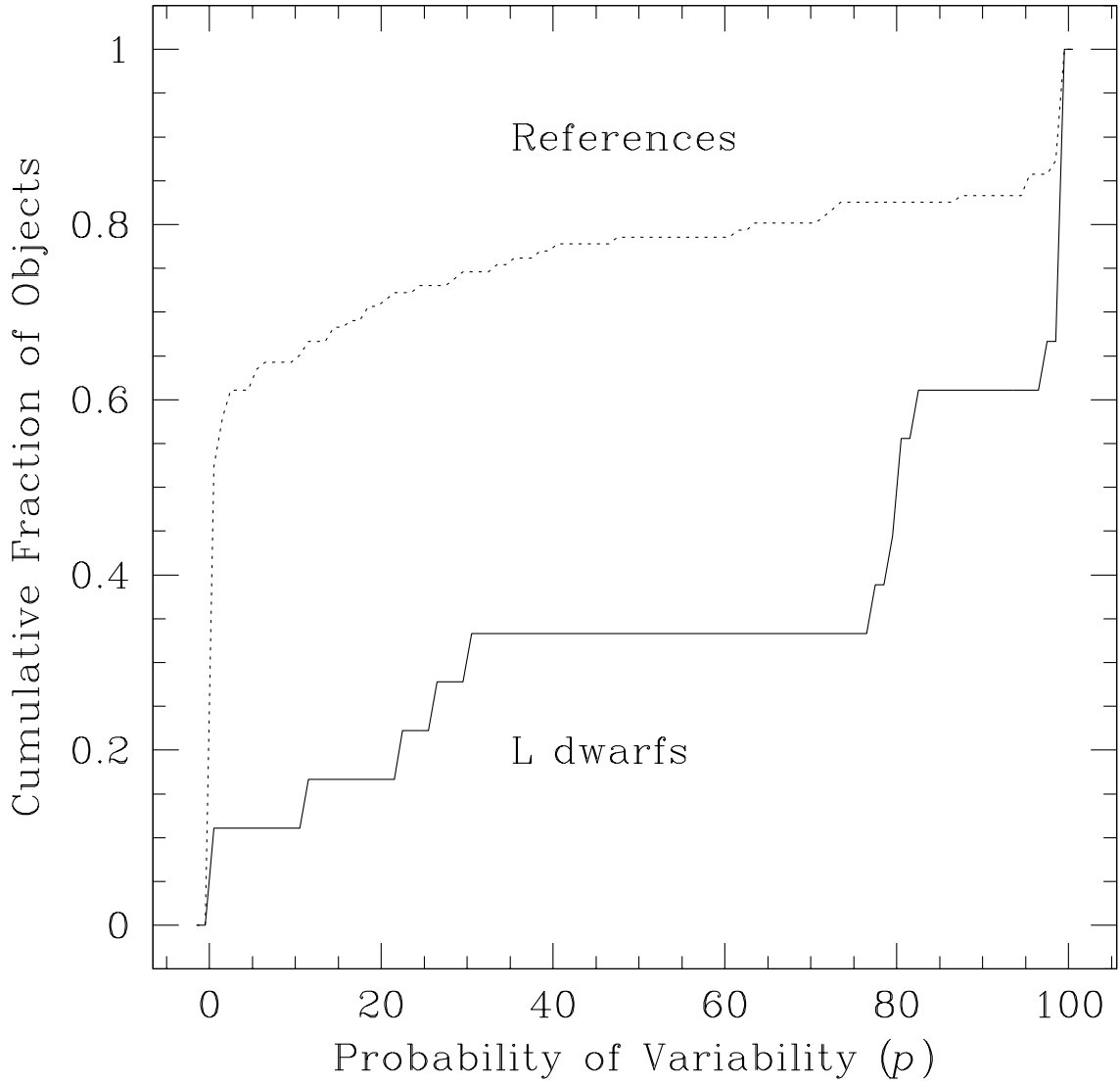}
\caption{Cumulative fraction of L dwarfs (solid line) and reference stars 
(dotted line) as a function of probability of variability.  The clear 
trend is that the L dwarfs tend to have higher values of $p$ than the
references, indicating that the L dwarfs are more variable. \label{perhist}}
\end{figure}

\begin{figure}
\plotone{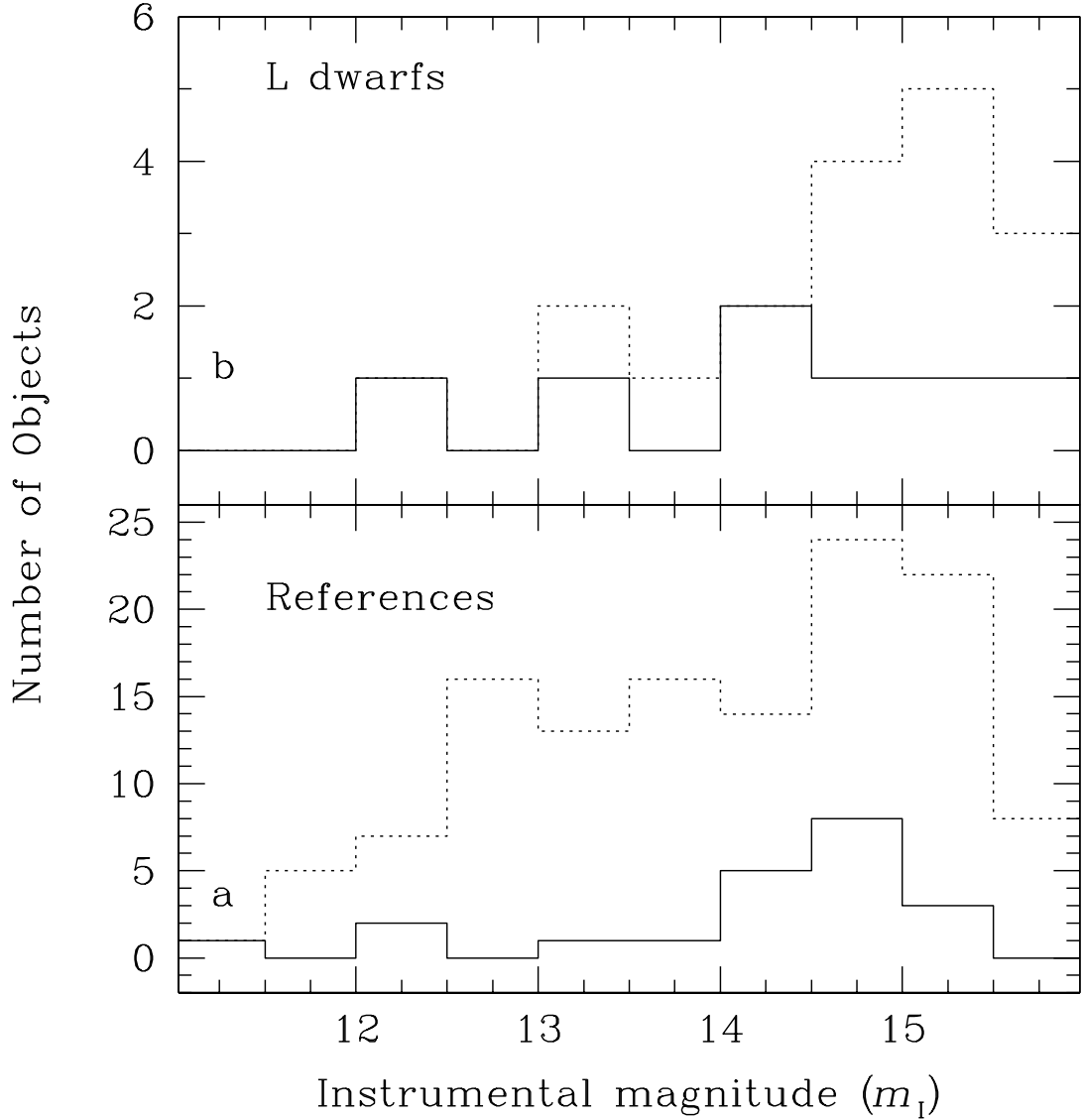}
\caption{a: Histograms of the numbers of variable (solid line) and total 
(dotted line) references as a function of magnitude.  The lack of any 
systematic trend indicates that we are adequately estimating the photometric
errors in these objects.  b: Same as panel a except for the L dwarfs.
  \label{histogram}}
\end{figure}

Supporting of our claim for variability in the L dwarfs, Figure~\ref{perhist}
shows the fraction of the
total number of objects with a given probability of variability. 
Over 60\% of the references have 
probabilities below 5\%, whereas only 10\% of the L dwarfs do.  In addition,
about 40\% of the L dwarfs are flagged as variable ($p>95.4$\%) compared to 
about 15\% for the references. Furthermore, Figure~\ref{histogram} shows the 
number of variable objects as a function of instrumental magnitude. The 
fraction
of references in a given magnitude bin that are variable does not show
any indication of being a strong function of object brightness, suggesting
that we are not underestimating the errors in these objects.  The slight
increase between $m_{\rm I}$=14-15 in the references is likely due to 
small number statistics.   Given this evidence, it seems clear that the L 
dwarfs derive from an inherently more variable population than the reference 
stars. 

Three of the targets in our sample were also monitored by \citet{bai01a}:
2MASS 1146+22AB, 2MASS 1439+19, and 2MASS 0345+12.  Our results disagree on 
the status of 2MASS 1146+22AB.  This object is significantly
non-variable in our sample, whereas \citet{bai01a} detected variability with
an RMS amplitude of 0.015 mag.  Our average photometric error for this object 
is almost four times larger than this amplitude.  
It is quite possible that 2MASS 1146+22AB varies with such a small amplitude 
as to be undetectable in this work.  Therefore, even though we find that this 
object does not display statistically
significant variations, higher signal-to-noise observations could show that it
is still a low-amplitude variable.  

The second object, 2MASS 1439+19, is 
classified as a possible variable here and a non-variable by Bailer-Jones \& 
Mundt.  Interestingly, their variability probability is larger than the one
we find here (90\% compared to 80\%).  Given that the probabilities are 
similarly high, this object is likely a low-amplitude variable.

\begin{figure}
\plotone{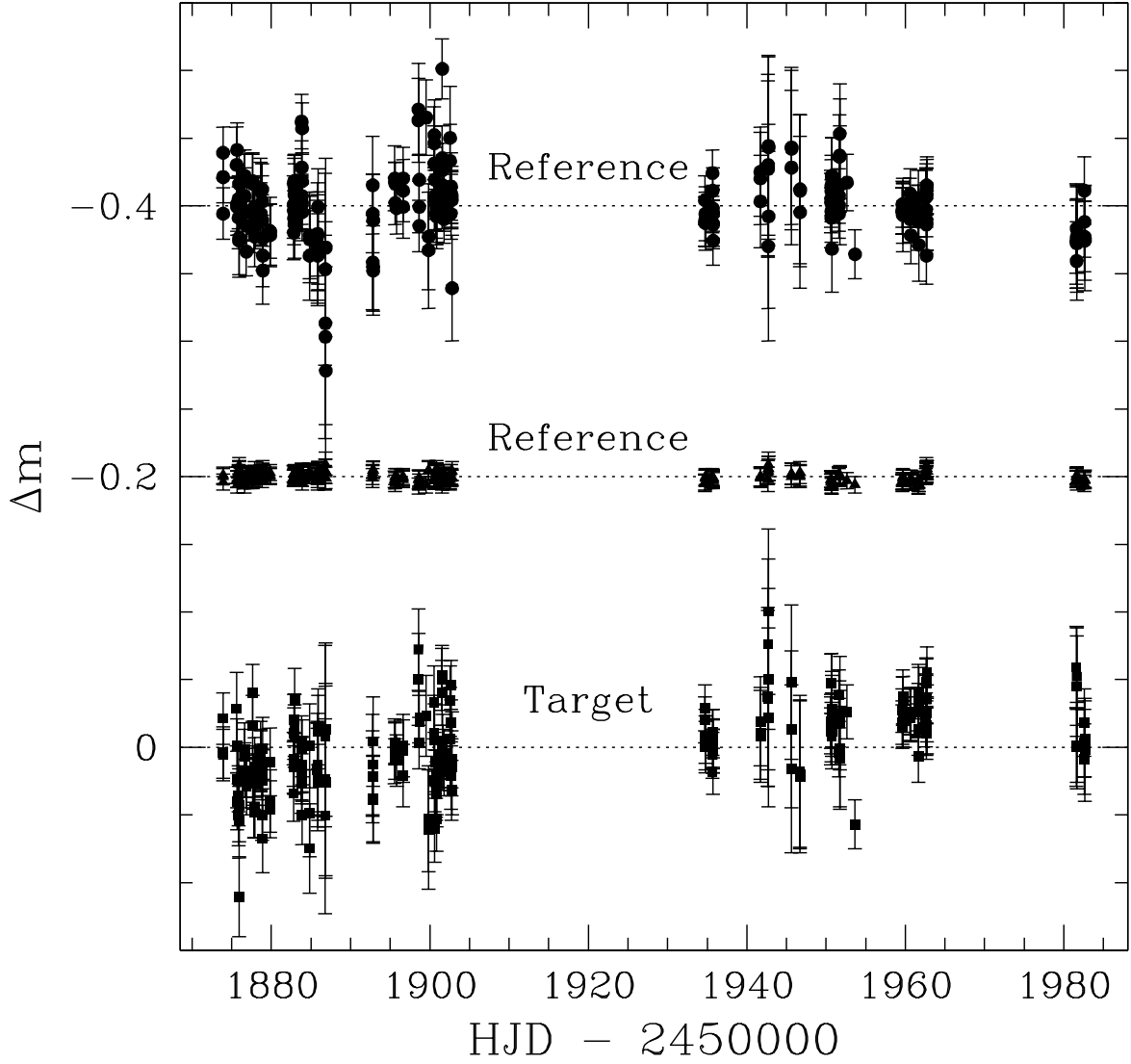}
\caption{Differential magnitude vs. HJD for 2MASS 0345+25 (squares), a bright
reference (triangles offset by $-$0.2 mag) and a faint reference (circles
offset by $-$0.4 mag).  Note that up (i.e. more negative $\Delta$m) represents
an increase in object brightness.
  \label{2m0345dif}}
\end{figure}

We agree with the results of \citet{bai01a} on the classification of 
2MASS 0345+12 as a variable object.  The light curve of 2MASS 0345+25 
(Figure~\ref{2m0345dif}) shows quite a bit of
scatter, but no periodic trends are evident.  The primary peak in the CLEAN
power spectrum occurs at 24.1$\pm$0.1 hours with a power $\approx$8 times 
higher than the noise.  A couple of reference stars
have prominent CLEAN peaks at 12 hours, indicating that the period is not 
intrinsic to any one object.  Consequently, we list 2MASS 0345+25 as a 
variable object in Table~\ref{tab-sv}, but do not present it in 
Table~\ref{tab-pow}.  

\begin{figure}
\plotone{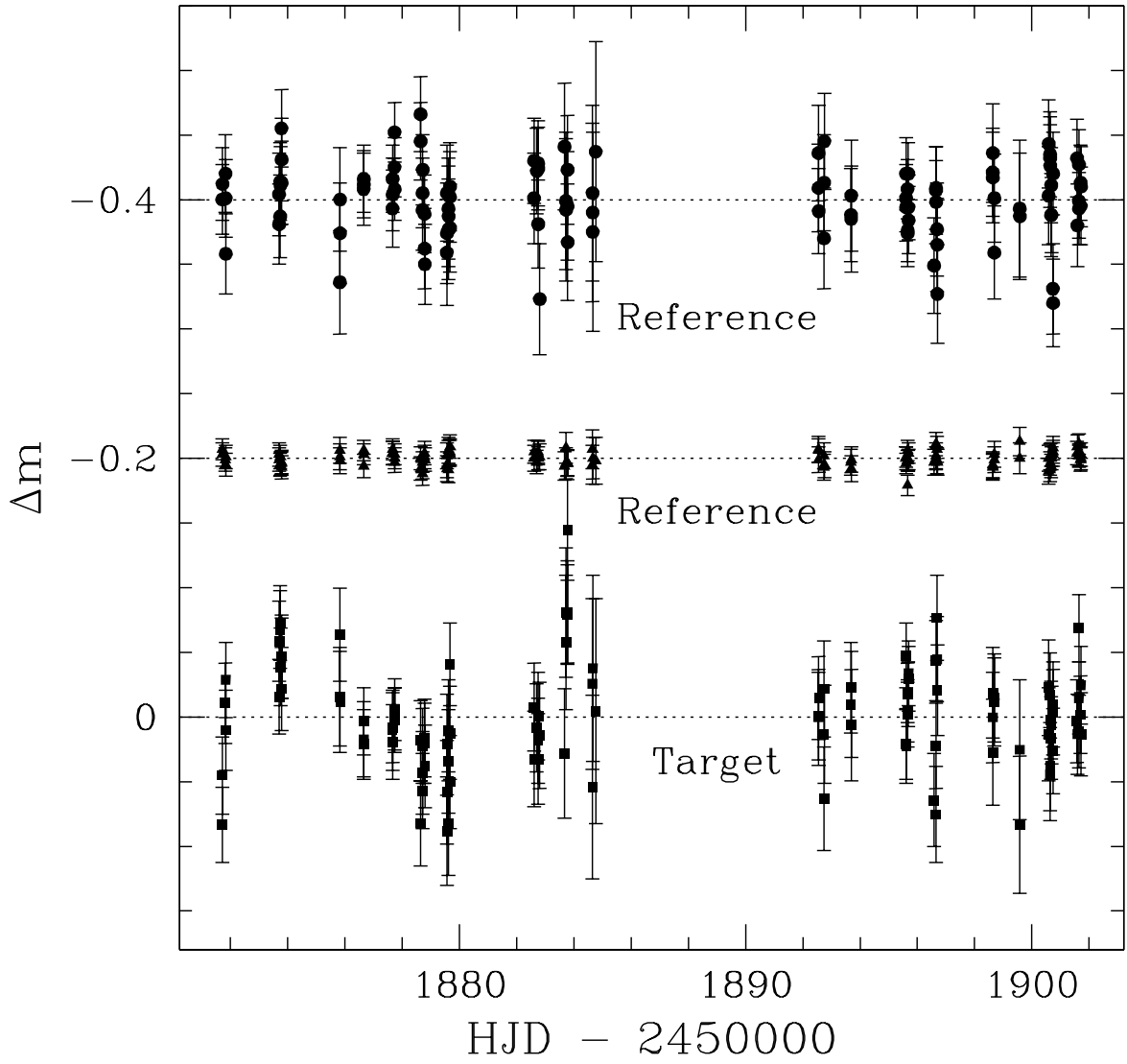}
\caption{Same as Figure~\ref{2m0345dif} except for 2MASS 0135+12.
  \label{2m0135dif}}
\end{figure}

\citet{gel01} presented preliminary results for three L dwarfs in this program.
They found 2MASS 0036+18 and 2MASS 0135+12 to be variable
and 2MASS 1412+16 to be non-variable. An error was found in the reduction 
process used to obtain those results.  The FWHM used to reject high FWHM 
frames was only from one star, and not an average of all stars.  Upon 
correction of the error, the classification of 
2MASS 0036+18 changed from variable to non-variable.  The classifications
of the other two objects did not change, but their data sets
did change slightly.  The new light curve for 2MASS 0135+12 is shown in 
(Figure~\ref{2m0135dif}).  This object does not have a significant period
(Table~\ref{tab-pow}), contrary to what was reported previously.

\begin{figure}
\plotone{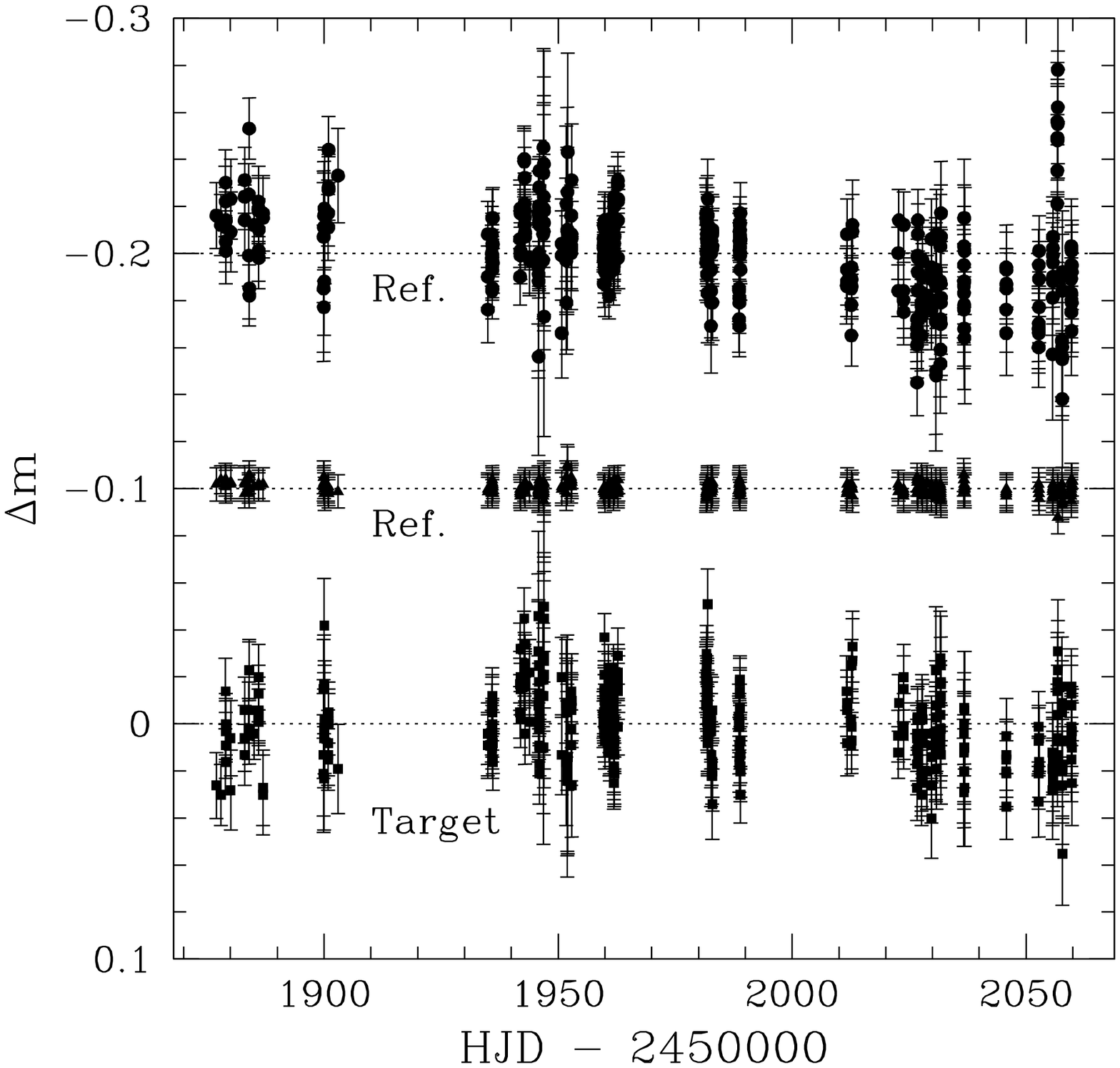}
\caption{Same as Figure~\ref{2m0345dif} except for 2MASS 1108+68 and the bright
and faint references are offset by $-$0.1 and $-$0.2 mag, respectively.
  \label{2m1108dif}}
\end{figure}

2MASS 1108+68 is another L dwarf displaying statistically significant 
variations and no significant period.
With a baseline of nearly 6 months (Figure~\ref{2m1108dif}), the data obtained 
for this object represents the most extensive 
photometric monitoring of any L dwarf to date.

\begin{figure}
\plotone{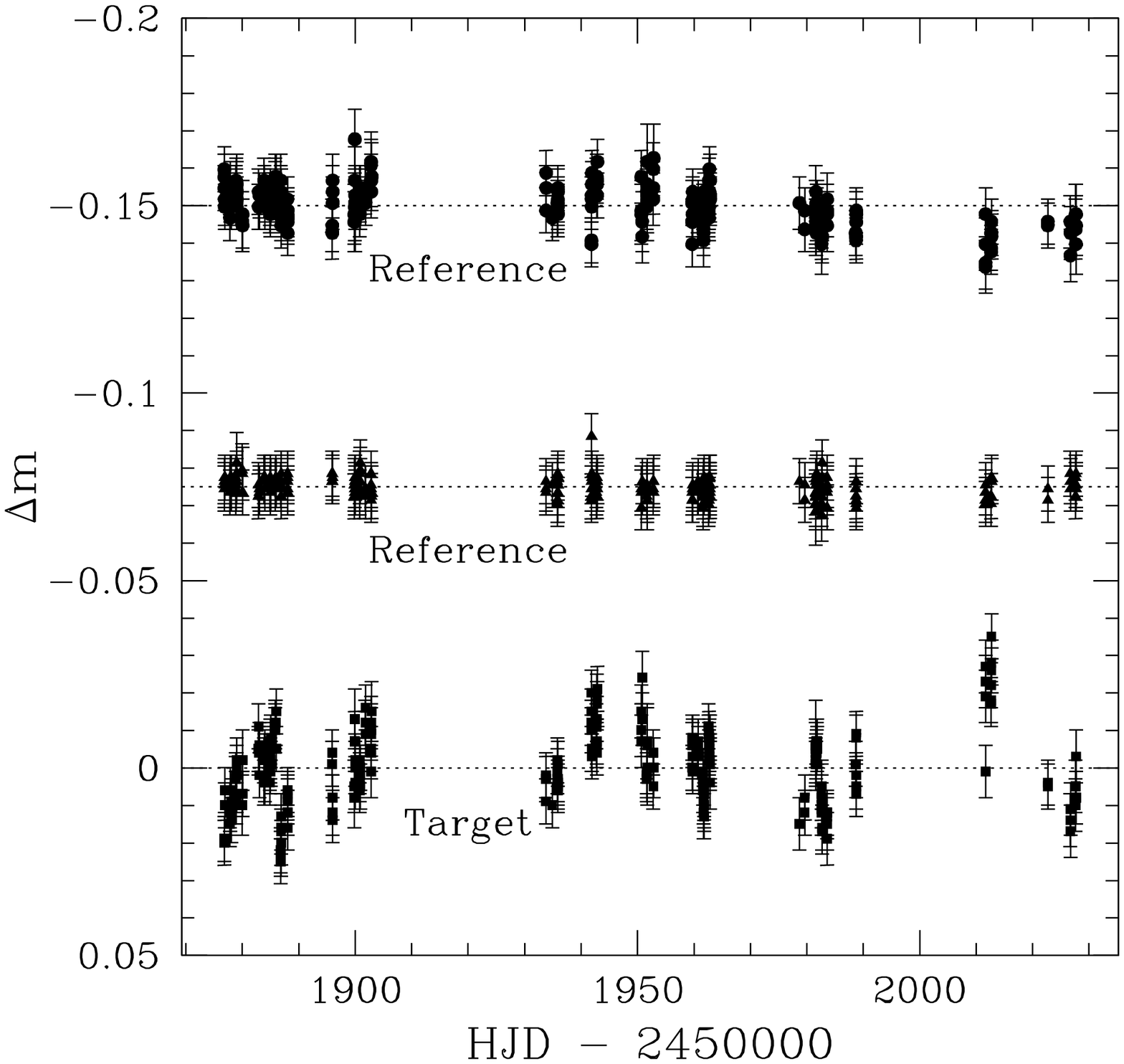}
\caption{Same as Figure~\ref{2m0345dif} except for 2MASS 0746+20AB and the 
bright and faint references are offset by $-$0.075 and $-$0.15 mag, 
respectively.
  \label{2m0746dif}}
\end{figure}

\begin{figure}
\plotone{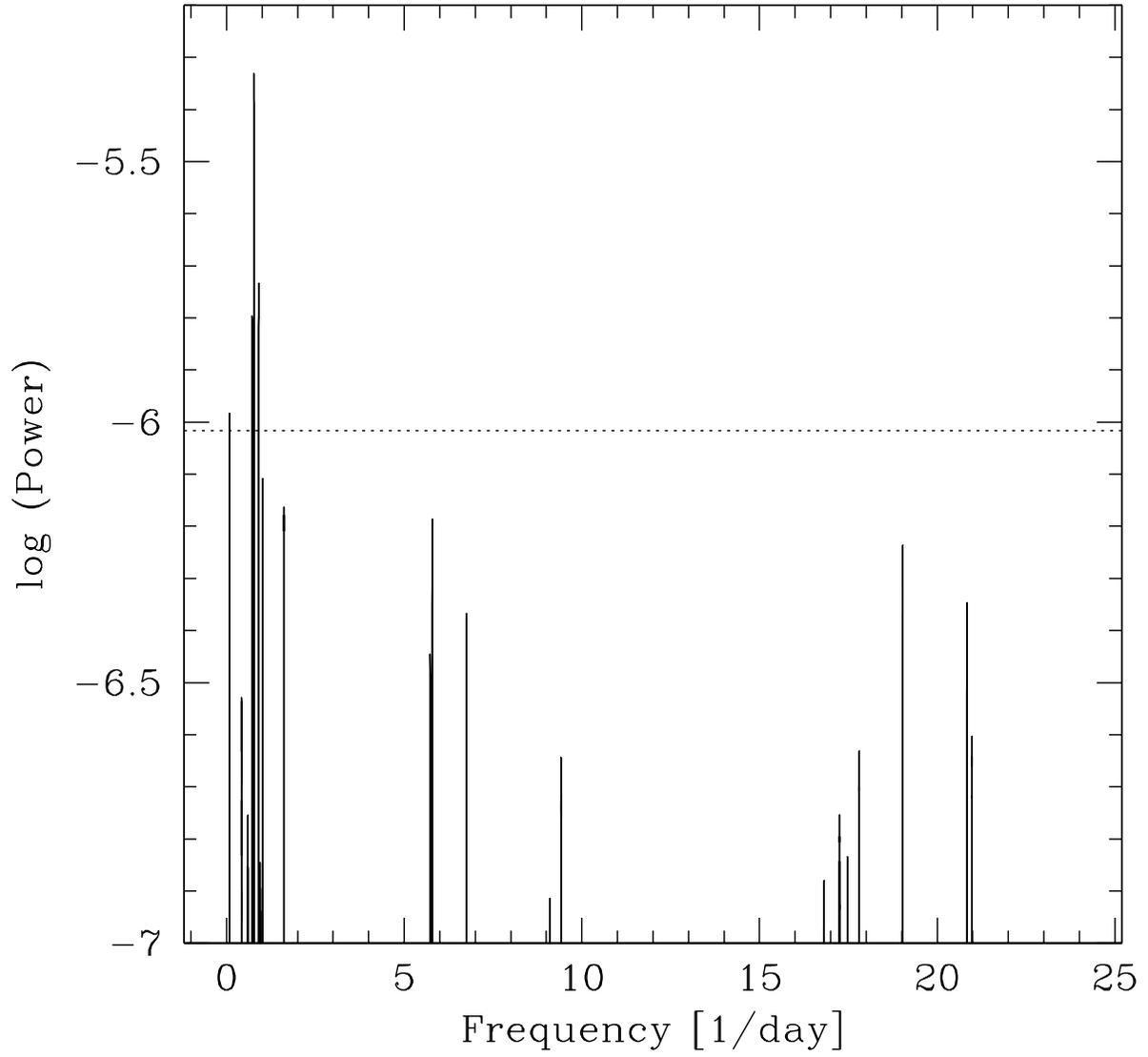}
\caption{Power spectrum of 2MASS 0746+20AB.  The dotted line denotes the noise 
level as defined by the average power of the 1000 random light curves.
The peak is located at 31.0$\pm$0.1 hours.
  \label{2m0746cln}}
\end{figure}

\begin{figure}
\plotone{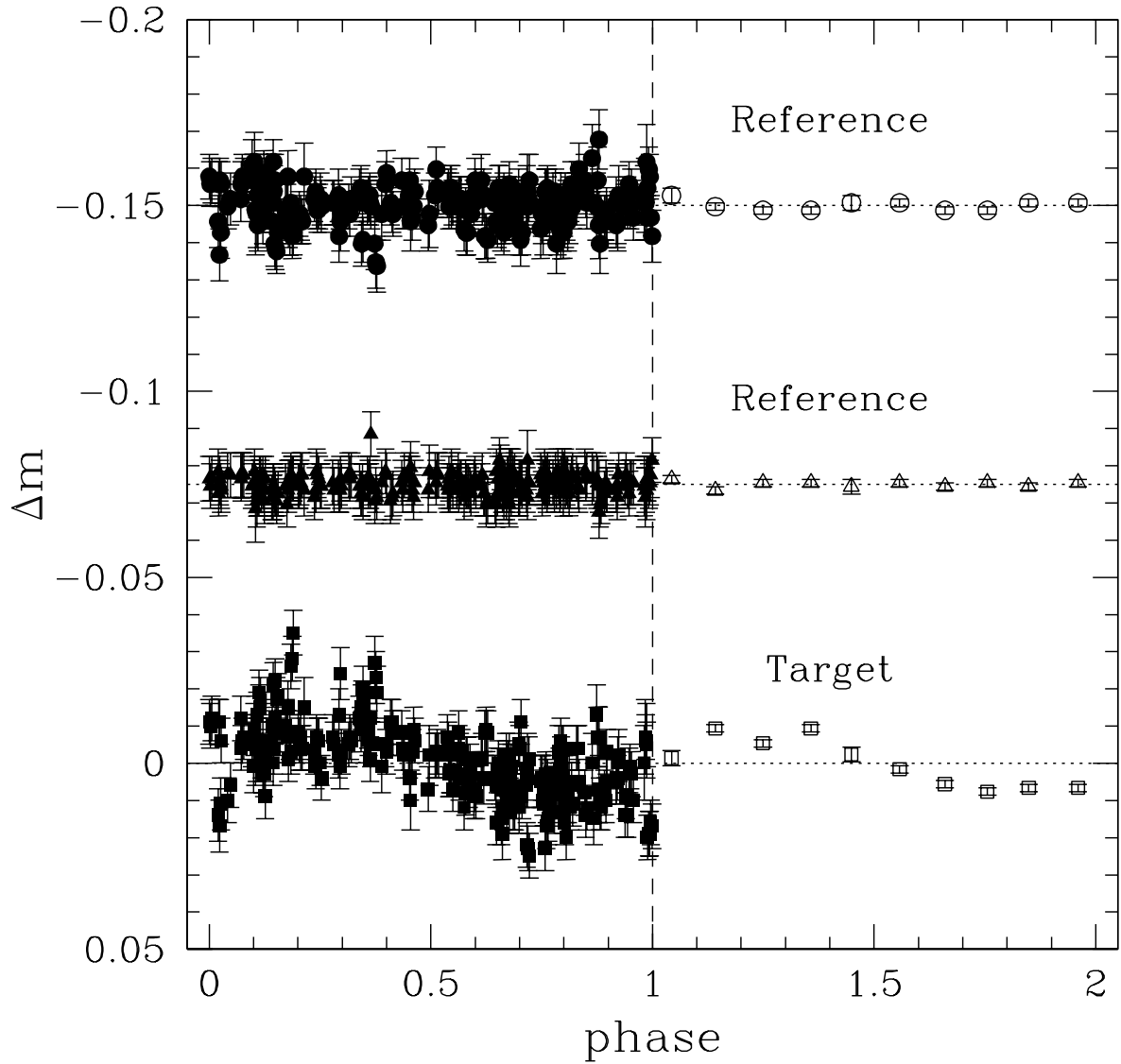}
\caption{Differential magnitude vs. phase for 2MASS 0746+20AB.  The points and 
offsets are the same as in Figure~\ref{2m0746dif} for phase = 0-1; the open
symbols from phase = 1-2 are averages of the data from bins 0.1 phase units
wide. 
  \label{2m0746phs}}
\end{figure}

Though not as extensive as 2MASS 1108+68, 2MASS 0746+20AB also has very good 
coverage (Figure~\ref{2m0746dif}).   This binary is the brightest object in our
target list and is variable.
The most dominant peak in the CLEAN power spectrum (Figure~\ref{2m0746cln})
is quite high 
compared to other peaks and is present at a period of 31.0$\pm$0.1 hours.  
The power of this period is nearly 5 times higher than the noise and the 
phased data (Figure~\ref{2m0746phs}) shows a roughly sinusoidal light curve.

\citet{bas00} present $v \sin i$ measurements for twelve L dwarfs, including
2MASS 1439+19 and 2MASS 1146+22AB from this study. The values they derive 
span from 10$\pm$2.5 to 60$\pm$5 km s$^{-1}$.  For typical brown dwarf radii 
\citep{bur97}, these velocities translate to rotation
periods $\lesssim$10 hours.  Because the inclination angle of 
the rotation axis is unknown, this period is an {\it upper} limit to the 
true rotation period of these objects.  Consequently, the 31-hour
period found here is likely too long to be related to the rotation period.

\begin{figure}
\plotone{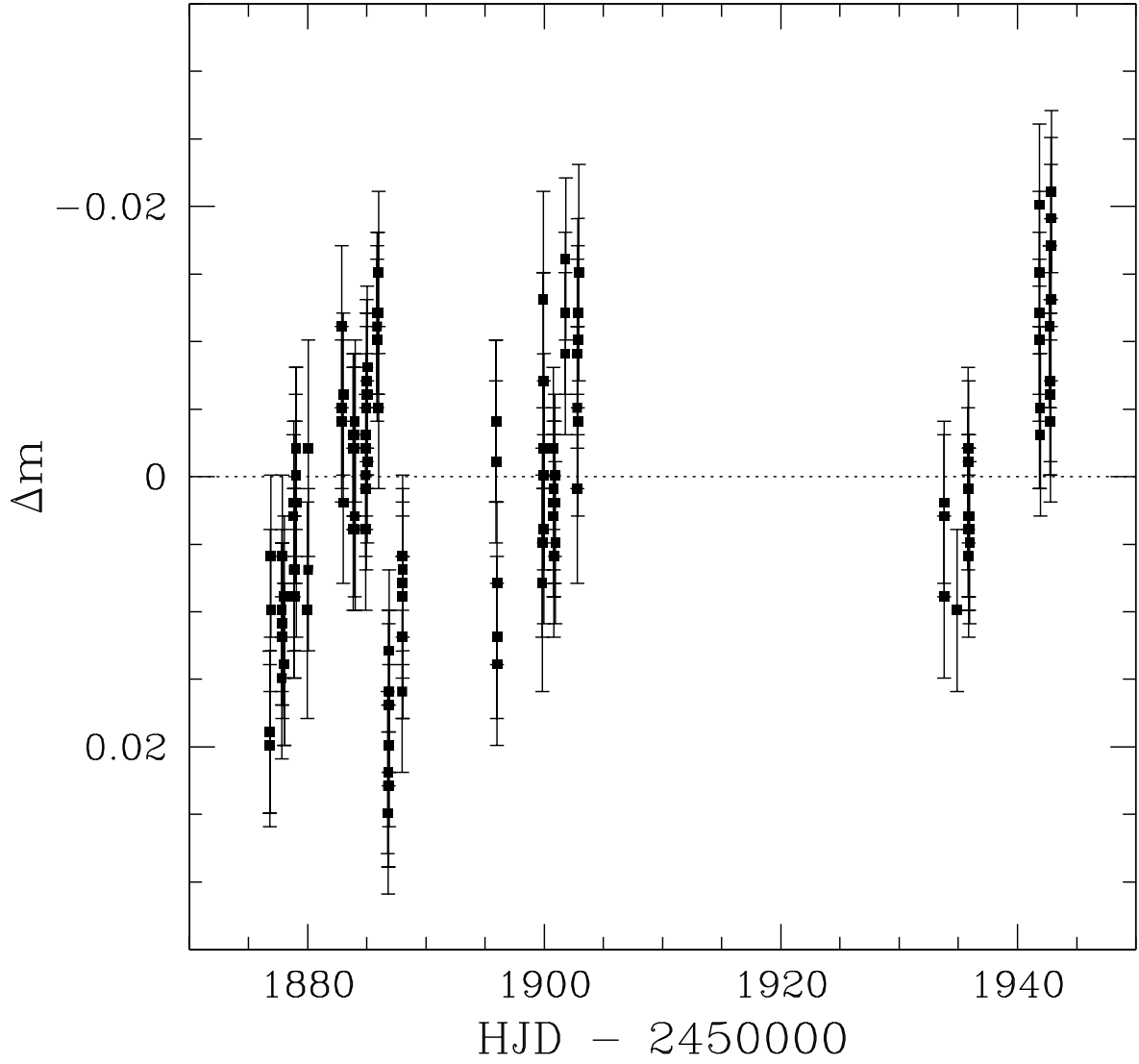}
\caption{Close-up view of the first 70 days in the light curve for 
2MASS 0746+20AB.  
  \label{2m0746zoom}}
\end{figure}

The first 70 days of coverage for 2MASS 0746+20AB reveal an interesting trend
(Figure~\ref{2m0746zoom}). 
The data appear to have a rough saw-tooth light curve that repeat with
a period of approximately 20 days.  After HJD 1950 this pattern is not as
prominent.  It is unclear what variability source can cause a saw-tooth pattern
such as this.  Spots and clouds should produce more gradual changes in the 
light curve.  A flare would produce a sudden brightening, followed by
a gradual dimming, the opposite to what is seen here.  Furthermore, 
\citet{reid01} estimate an orbital semi-major axis of 3.4 AU, giving a period
of about 18 years.  Therefore, both the shape and duration of the feature
are not what would be expected for an eclipsing system.

\begin{figure}
\plotone{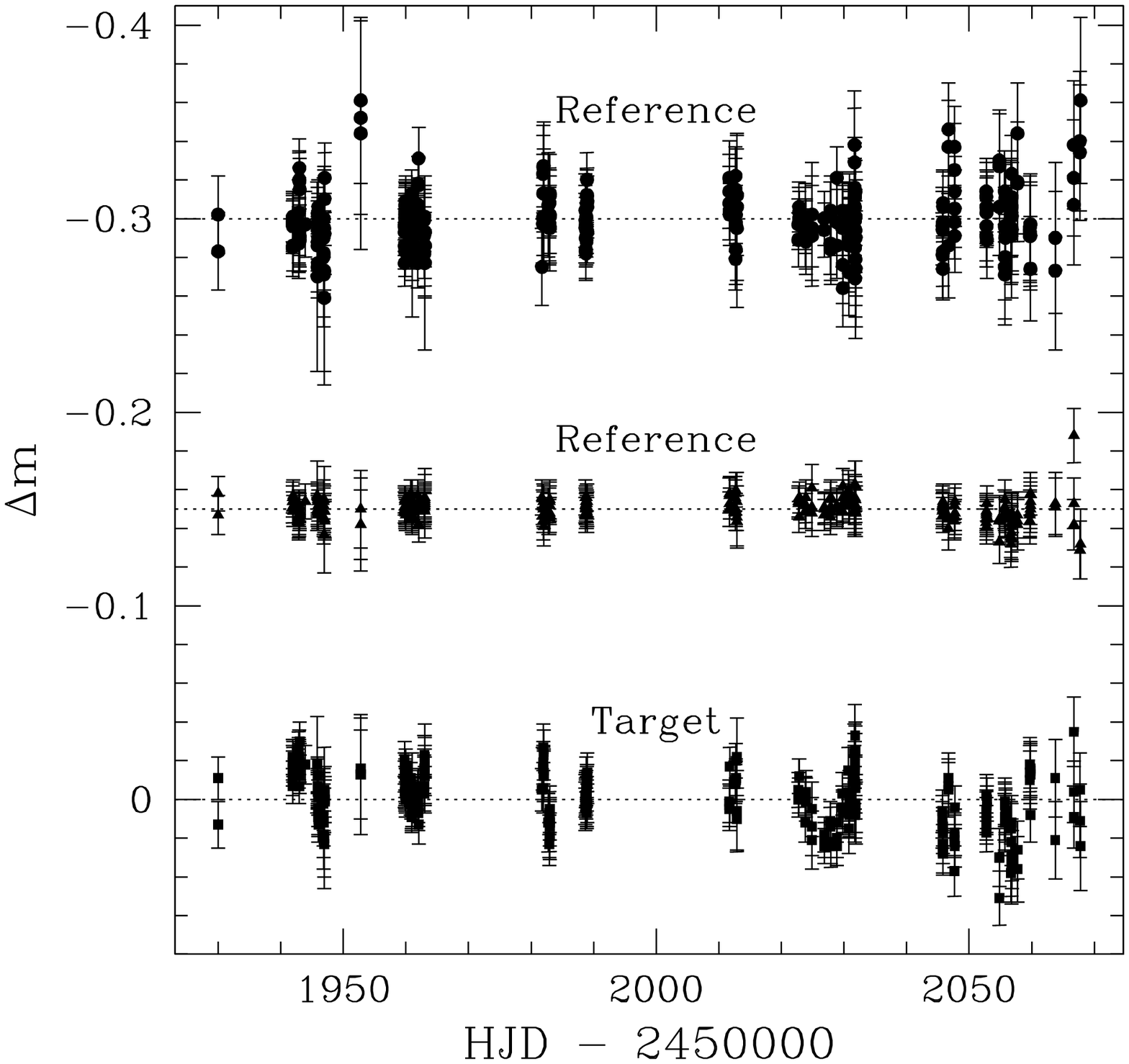}
\caption{Same as Figure~\ref{2m0345dif} except for 2MASS 1300+19 and the bright
and faint references are offset by $-$0.15 and $-$0.3 mag, respectively.
  \label{2m1300dif}}
\end{figure}

\begin{figure}
\plotone{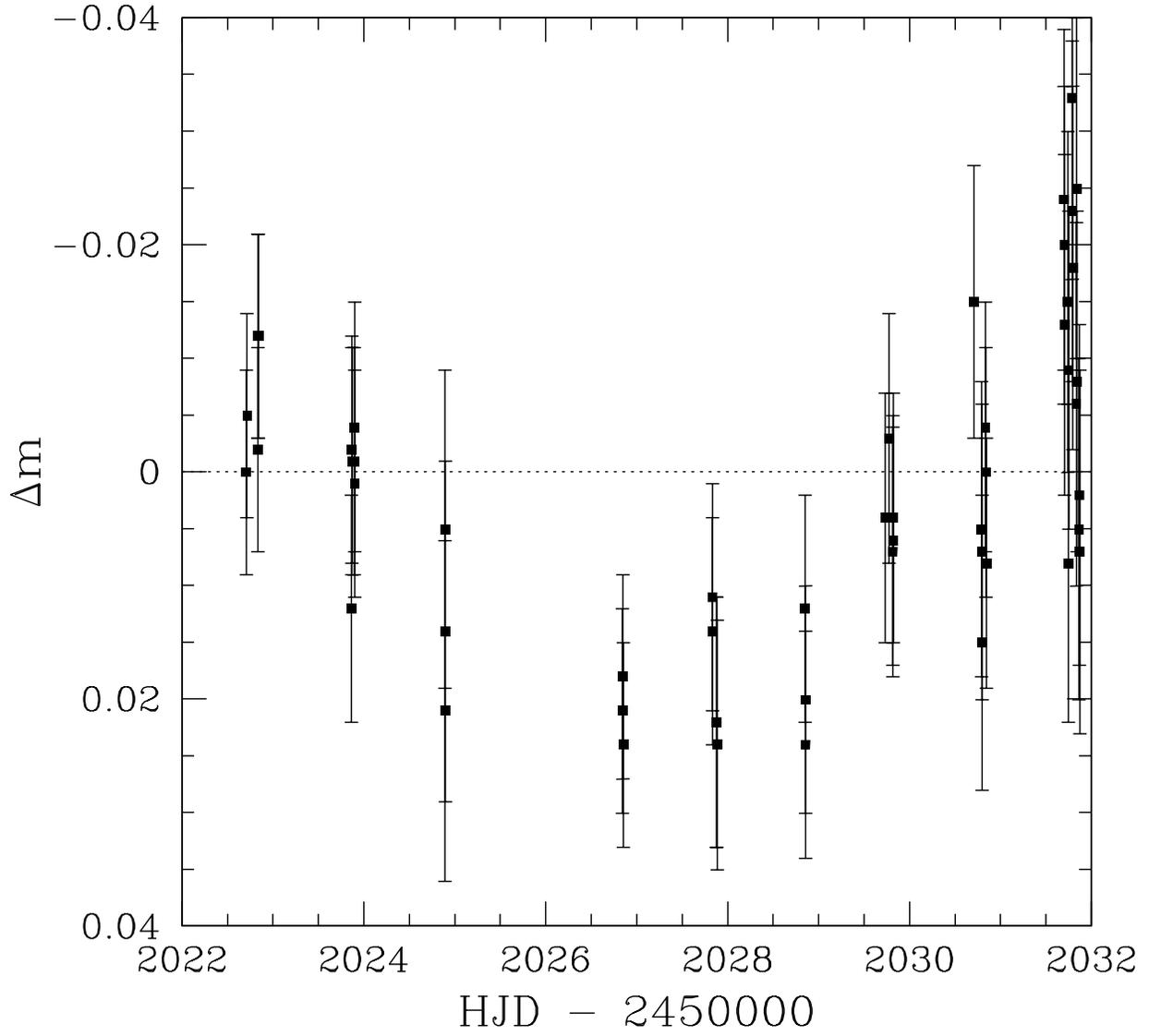}
\caption{Close-up view of the light curve around HJD=2027 for 2MASS 1300+19.  
  \label{2m1300zoom}}
\end{figure}

2MASS 1300+19 also has an interesting feature in its light curve
(Figure~\ref{2m1300dif}).
The sequence of points around HJD=2027 (Figure~\ref{2m1300zoom}) show roughly 
sinusoidal variations.  This span of points is approximately the same as the 
best period for the entire data set, 238 hours, indicating that this feature
is the source of the period.  The period is much longer than expected for a
rotation period.  It seems likely that some other cause is responsible for the
feature.

Although the feature is reminiscent of an eclipsing binary light curve, 
this scenario seems unlikely.  The depth of the 
feature implies a secondary object radius about 18\% the radius of the L dwarf,
roughly 3 times larger than Earth.  The duration suggests an orbital separation
$>500$ AU.  The probability that we should observe an edge-on binary system 
with an orbital separation $>500$ AU just as it is eclipsing is extremely small
($< 10^{-7}$).  Furthermore, while there are stars with L and T dwarf 
companions at separations
greater than 500 AU, there are no known companions to L dwarfs beyond 10 AU.

It is possible that the 10-day feature in this object's light curve is
related only to its surface features.  The sudden creation and dissipation of
a large storm could possibly produce the changes we see.  A similar event 
happened with Saturn several years ago \citep{beebe92}, but over a longer 
dissipation
timescale.  However, features on Jupiter and Saturn are known to evolve on
timescales from hours to years, depending on the features' positions and 
rotation orientations (Beebe 2001, private communication), so the timescale
seen here is certainly plausible.
A large storm would not only change the brightness of the object, it should
also have an effect on the photometric colors.  

Table~\ref{tab-pow} presents the primary CLEAN periods and their PNR for
six variable and five possibly variable L dwarfs.  The only periods 
considered significant are those with PNR
values above 3, 2MASS 0746+20AB and 2MASS 1300+19.  We have discussed 
why we do not believe these periods to be related to the 
rotation period of these objects.  

In addition to the objects mentioned above, our sample contains several
other objects (Table~\ref{tab-sv}).  These objects do not have
significant periods, nor do they have any interesting features in their light 
curves.  The light curves for all objects will be found in \citet{gel02}.

\section{Possible Variability Sources}
We have discussed the existence of statistically significant photometric 
variations in seven of eighteen L dwarfs and possible variations in another
five. These variable
objects cover the entire span of spectral types present in our sample, from 
L0 to L4.5.  Their data were sampled with a minimum timescale of $\sim$5 
minutes, intra-night sampling $\sim$1-2 hours, and maximum baselines between 26
and 230 days.  Most of these variables show no significant 
periodicity.   Whatever the source of 
these variations, it must be present in a variety of spectral
types and, hence, effective temperatures. 

Two properties are commonly mentioned in the literature in which L 
dwarfs can be unusual: H$\alpha$ and variability. These may or may not be 
related since some of 
the variable L dwarfs have H$\alpha$ present in their spectra and
others do not.  We now discuss possible sources for the variations and their 
likelihood of being the present in these objects.

\subsection{Magnetic Spots}
\subsubsection{Summary of Magnetic Activity Observations}
\begin{figure}
\plotone{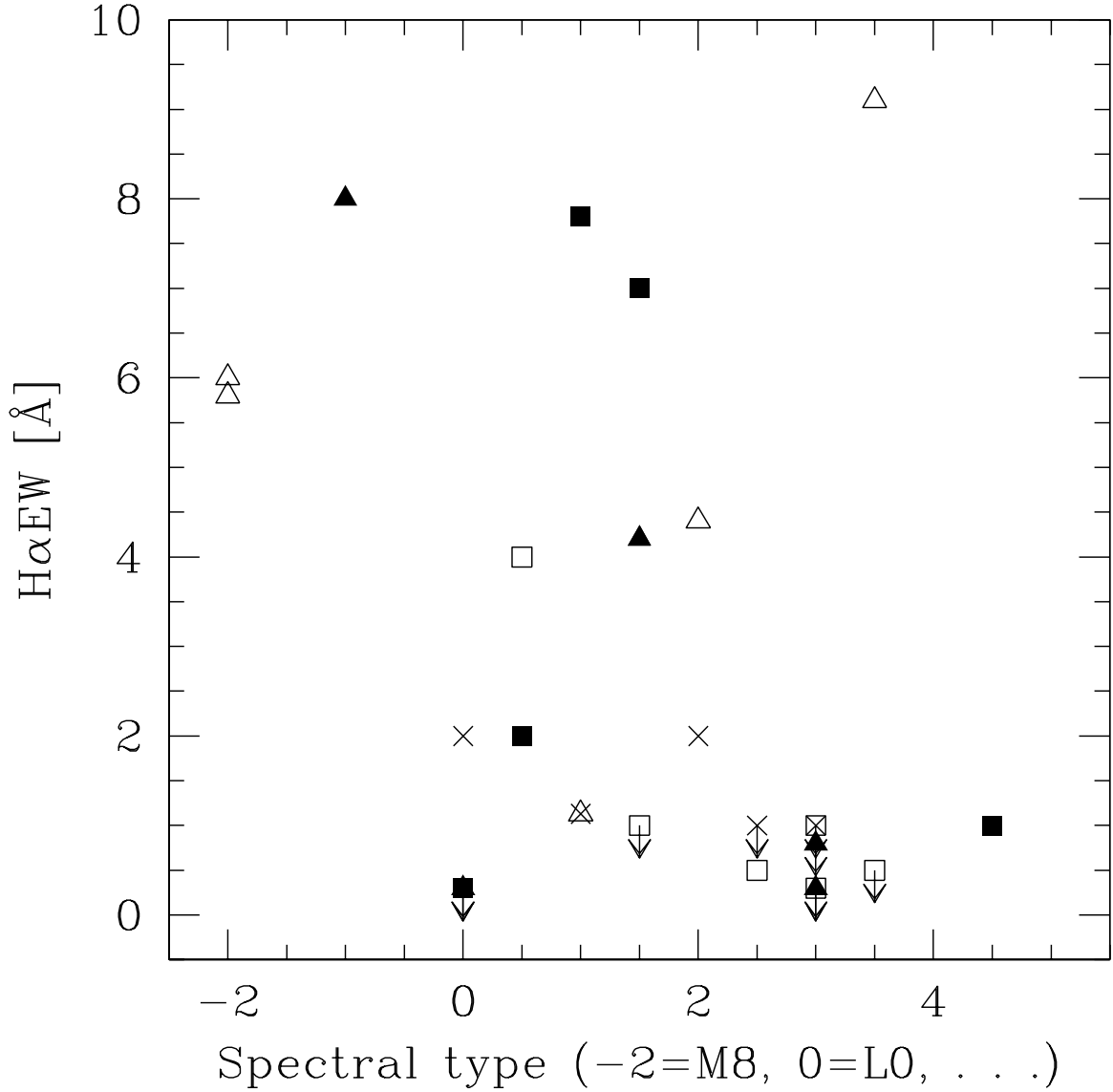}
\caption{H$\alpha$ equivalent width as a function of spectral type for 
variables (solid symbols), non-variables (open symbols), and possible
variables (crosses) from this study (squares) and \citep{bai01a} (triangles).
H$\alpha$ upper limits are shown with arrows.  The lack of any trends between
variability and H$\alpha$ emission suggests that magnetic activity is not 
responsible for photometric variability.
  \label{hafig}}
\end{figure}

Many L dwarfs exhibit 
H$\alpha$ emission \citep{kir99,kir00}, which is known to be an 
indicator of high chromospheric temperatures and magnetic 
activity in earlier type stars \citep{haw00}.  If the variability
in L dwarfs is caused by magnetic spots, then it is plausible to
expect a correlation between H$\alpha$ emission and the variable
objects.  In agreement with the conclusions of \citep{bai01a} and  
\citep{mart01}, Figure~\ref{hafig} shows no 
correlation between H$\alpha$ emission (i.e. magnetic activity) and
variability for the sample of L dwarfs from \citep{bai01a} and this study.  
This result could imply that either magnetic activity
is not the source of the variation or that the H$\alpha$ emission is
not magnetic in origin for these objects.  

As with earlier-type stars, wave heating has been suggested as
the mechanism responsible for producing hot brown dwarf upper
atmospheres \citep{yel00}. Convection-forced waves propagate
and grow as they rise through the upper atmosphere, eventually 
releasing their energy and heating the gas as they dissipate. 
The resulting high temperatures
combined with magnetic field effects are possibly responsible for
the  H$\alpha$ emission.  L dwarfs with weak magnetic fields should
exhibit little or no H$\alpha$ emission.  In light of this it is
quite notable that the fraction of objects with H$\alpha$ emission
peaks at spectral type M7 and decreases at earlier and later spectral
types \citep{giz00}.  No L dwarfs later than L5 show
H$\alpha$ in emission \citep{kir01}, although one T dwarf does 
\citep{burg00}.  Since the early L population consists of both
young brown dwarfs and old stars, this trend could indicate an inability
of either substellar objects or cool stars to produce
magnetic fields appropriate to maintain the H$\alpha$ 
emission \citep{giz00}.

Using kinematics as a probe for age, \citet{giz00} have argued that the
substellar late-type dwarfs show less H$\alpha$ emission than 
stellar dwarfs of the same effective temperature.  This might imply that
the process by which H$\alpha$ emission is produced is driven by the
mass of the object and not the effective temperature.  Indeed, the
dissipation of acoustic waves in the upper atmosphere could be
responsible for the heating of the chromosphere.  Unfortunately, this
process is poorly characterized at the masses and effective temperatures
of interest here.

In addition to H$\alpha$ emission, radio emission can also be a
signature of a magnetic field.  Radio flares as well as quiescent emission
have recently been reported for the M-dwarfs LP944$-$20 \citep{ber01} and 
BRI0021 \citep{berger02}, and the L3.5 dwarf 2MASS 0036+18 \citep{berger02}, 
an object in our sample.
These authors infer that the radio emission is caused by synchrotron emission
and estimate field strengths of $\sim$5 G for LP944$-$20, 5-50 G for BRI0021 
and 20-350 G for 2MASS 0036+18.  Since there are essentially no methods
to accurately determine the magnetic field strengths of objects other
than the sun \citep*{hai91}, these values are based upon models and
assumptions that might not be correct.  Regardless of these assumptions,
however, \citet{berger02} surmises that this data imply a substantial,
non-neutral corona in this L dwarf, although the mechanics to create
such a corona are poorly understood.    

If the magnetic field strengths are correct for LP944$-$20, BRI0021,
and 2MASS 0036+18, they are apparently much less than the field strengths of 
active M dwarfs \citep{hai91}.  In addition, the substellar
nature, old age \citep{tin98a}, and rapid rotation \citep{tin98b} of 
LP944$-$20 support a weak field strength; many L dwarfs with
spectroscopically determined rotation velocities are rotating quite
rapidly and lack significant H$\alpha$ emission \citep{haw00},
suggesting that the magnetic fields of these presumably old objects are
too weak to slow down the rotation.  The rotation velocity of
2MASS 0036+18 is unknown and it has no measurable H$\alpha$ emission.

The existence of a magnetic field in LP944$-$20 is also supported by the
observation of an X-ray flare \citep{rut00}.  An X-ray flare in an old,
non-accreting object such as this can only be caused by magnetic
activity.  However, the lack of quiescent X-ray emission suggests
that the magnetic field is quite weak.  Rutledge et al. postulate
that because of the rapid rotation in this object, either the turbulent dynamo
is being suppressed or the magnetic field is being configured into a
more organized form.  They also suggest that the lack of ionization in
the cool photosphere prevents the magnetic field from coupling with the
gas in the atmosphere, causing the field to dissipate.

\citet*{fle00} arrive at the same conclusion with their study of the
X-ray flare of the M8 dwarf VB 10.  They go so far as to estimate the
ionization fraction in its atmosphere, by extrapolating the ionization
fractions in the atmospheres of early dwarfs down to the effective
temperature of VB 10.  They estimate that the ionization fraction in
late M dwarfs is 2 orders of magnitude lower than in early M
dwarfs and 3 orders of magnitude smaller than in the sun, and conclude
that magnetic footprints (i.e. spots) are unlikely to exist in 
the photosphere.  The cooler M9 dwarf LP944$-$20
is even less likely to have magnetic spots than VB 10, supporting
 the conclusion by \citet{tin99} that they were
detecting the signature of clouds.  This, in turn, implies that
magnetically produced spots are unlikely to be found in L dwarfs.
However, it is still useful to examine what conditions are needed for
the formation of magnetic spots and to determine if
these conditions exist in L dwarfs.

\subsubsection{Model Predictions}
Magnetic spots in the sun (i.e. sunspots) are thought to form
as magnetic flux tubes rise to the photosphere \citep{par55}.  For
magnetic buoyancy to be important, the plasma must be a sufficiently
good conductor.  This criterion is most likely not satisfied in
cool L-dwarf atmospheres, especially in the low pressure regions where
the temperatures are also low.  
In these regions the free electron abundance is small, $\sim10^{11}$ cm$^{-3}$
around 1 bar for a 2000 K model.  While this density is only an order of
magnitude less than coronal densities estimated by \citet{berger02}, it is
a factor of 10$^7$ smaller than the densities of neutral species (e.g. H$_2$,
and He) at this atmospheric pressure.  The corona is likely a low-density
region populated primarily by free electrons, whereas the atmosphere at a
pressure of 1 bar is largely neutral.  Therefore, these two regions should 
have substantially different electrical conductivities. The small fraction 
of free electrons relative to neutral species in the atmosphere suggests 
that this region should not be a good conductor, nor should it be able to 
support the formation of any magnetic spots. Nonetheless, it is instructive 
to explore the possibility of magnetic spots in more detail.

To estimate the strength of coupling between the gas and the magnetic
field we compute the magnetic Reynolds number $R_m$ for L-dwarf 
atmospheres.  $R_m$ is a dimensionless parameter
describing how efficiently a gas interacts with a magnetic field;
$R_m=lv/\eta$ \citep{pri82}, where $l$ is a length scale, $v$ is
a velocity scale,
and $\eta$ is the magnetic diffusivity of the gas.  When $R_m\ll$~1, 
the magnetic field slips through the gas with no interaction; for
$R_m\gg$~1, the magnetic field is frozen in the gas.

To compute $R_m$ we rely on atmosphere models computed by
\citet{marl02} for cloudy L dwarfs.  The models employ the
cloud model of \citet{ack01} with the best-fitting sedimentation parameter
$f_{\rm rain} = 3$.  The L-dwarf $T_{\rm eff}$ range is still uncertain,
but likely lies between about 2200 and 1300 K.  We consider models with
$T_{\rm eff}$ of 2000 to 1200 K and surface
gravity of $10^5$ cm s$^{-2}$, appropriate for a $\sim 35$ Jupiter
mass brown dwarf.

The appropriate length scale $l$ to use in the calculation of
$R_m$ is not obvious.  In the area around sunspots, $l$ is usually
taken as the size of the sunspot, a small fraction of the
solar radius.  We choose to set $l=H$, the pressure scale height.
Typical values of $H$ are around 10$^6$ cm
at the 1 bar level. For the velocity scale we use that predicted by mixing 
length theory.  To place a conservative upper limit on $v$ and
hence $R_m$, we assume that the entire thermal flux of the L dwarf is 
carried by convection.  This is reasonably accurate below the photosphere, but 
overestimates the velocity scale, and $R_m$, above.
Typical convective velocities are computed to be 10$^3$-10$^4$ cm s$^{-1}$.

The value for $\eta$ is computed from \citet{pri82};
\begin{equation}
\eta = 5.2 \times 10^{11}~\ln \Lambda~T^{-3/2}~A~{\rm cm^2\,s^{-1}},
\end{equation}
where
\begin{equation}
A \approx 1 + 5.2 \times 10^{-11} \frac{n_n}{n_e} \frac{T^2}{\ln \Lambda}
\end{equation}
is a factor to account for the partial ionization of the plasma, $T$
is temperature, $n_n$ is the number density of neutral
atoms and molecules, $n_e$ is the number density of electrons, and
\begin{equation}
\label{eq:coulomb}
\ln \Lambda \approx \ln \biggl(1.24\times10^4 \frac{T^{3/2}}{n_e^{1/2}} \biggr)~~{\rm for}~T<5.8\times10^5~ {\rm K}
\end{equation}
is the Coulomb logarithm \citep{somov92}.
We use the abundance tables of \citet{lod99} to calculate $n_e$ and
$n_n$.  At a pressure of 1 bar $n_n$ is $\sim10^{18}$ cm$^{-3}$ for all models,
whereas values of $n_e$ at this level are $\sim10^7$ cm$^{-3}$ for the 1200 K
model and $\sim10^{11}$ cm$^{-3}$ for the 2000 K model. 

Equation~\ref{eq:coulomb} provides a useful relation
for computing the Coulomb logarithm at temperatures below 10$^6$ K. 
This relation might not be valid in the atmospheres of L dwarfs, where
the temperatures of interest here do not exceed 4100 K.  For a given
model, $\ln \Lambda$ as calculated above varies between 5 and 30 
throughout the atmosphere.  In order to test the validity of these
results, we computed $R_m$ with $\ln \Lambda$ at the
unrealistic values of 1 and 100. We find that $R_m$ computed at these limits 
differs by less than 1 part in $10^6$ for
our coolest model and $<1\%$ for our warmest model and stress
that for the region of parameter space covered by our calculation
the precise value of $\ln \Lambda$ is essentially irrelevant.
Nonetheless, for the results considered here, $\ln \Lambda$ is computed
with Equation~\ref{eq:coulomb}.

\begin{figure}
\plotone{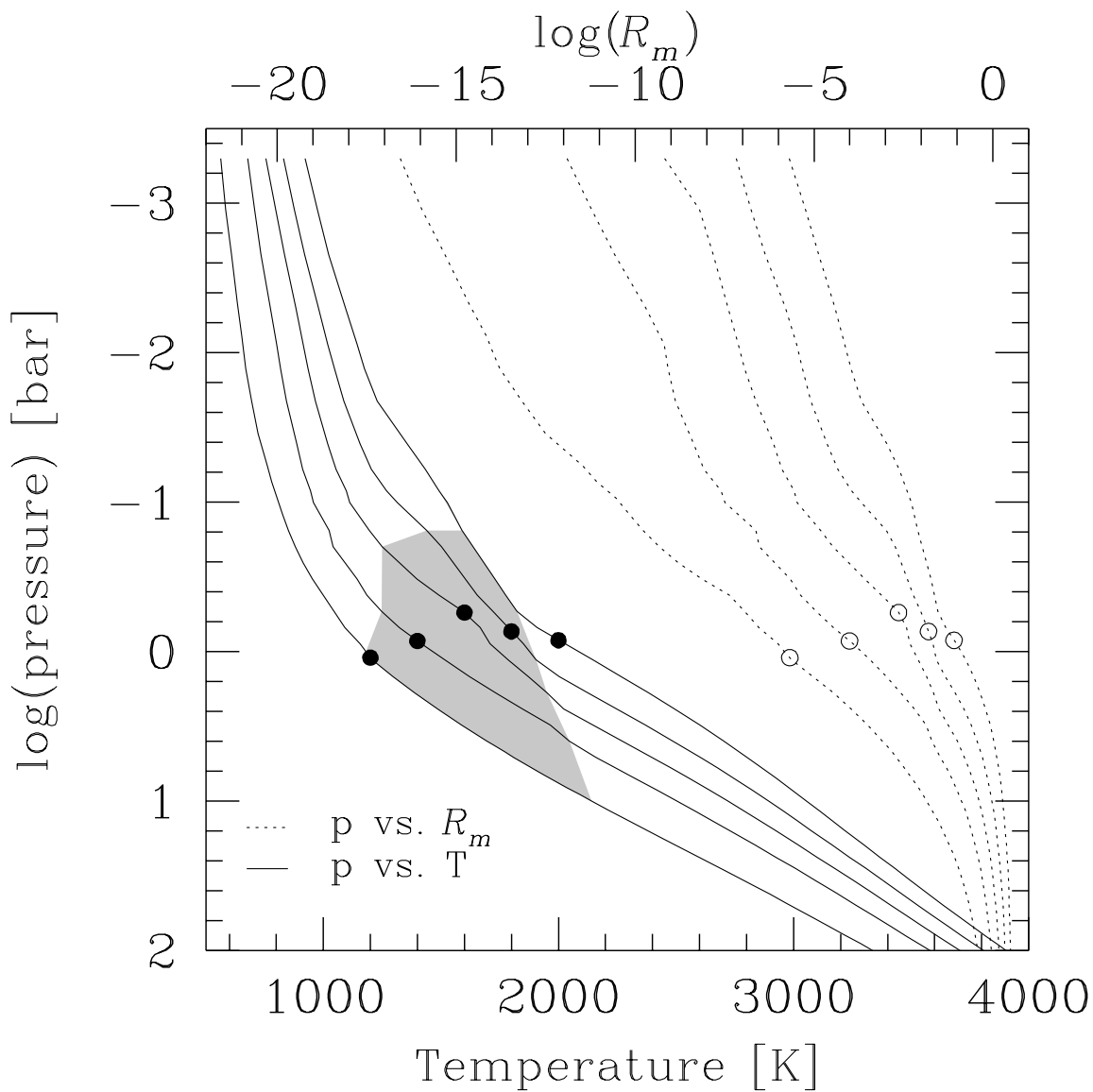}
\caption{Magnetic Reynolds number ($R_m$, dotted lines) and temperature
(solid lines) plotted as a function of pressure in the atmospheres of
L-dwarf models with $f_{\rm rain}$ = 3 and $T_{\rm eff}$ = 1200, 1400, 1600, 
1800, \& 2000 K (spectral types T1 to L2; Stephens et al. 2001),
going from left to right.  The solid dots represent the pressure at which
the temperature matches $T_{\rm eff}$ (approximately the photosphere);
open circles are the photosphere location in p vs. $R_m$ space;
the shaded area is the
region from the cloud bottom to cloud top (defined here as the level where
the cloud's integrated optical depth is $\sim0.1$).  $R_m$ is quite small 
throughout the entire atmosphere and starts approaching 1 below the base of 
the clouds and the level of the photosphere. 
\label{fig-rey}}
\end{figure}

The value of $R_m$ as a function of pressure in the model atmospheres is
shown in Figure~\ref{fig-rey}.  $R_m$ is very small throughout the entire upper
atmosphere; only at pressures of $\sim 100\,\rm bar$ and higher does
$R_m$ start to approach 1.  By comparison, $R_m$ near sunspots is
estimated to
be 10$^4$-10$^6$ \citep{pri82}.  Open circles denote the approximate
base of the photosphere (where $T=T_{\rm eff}$) for the models shown.  Note 
that in contrast to the sun, $R_m$ only approaches unity well below the
photosphere.
At $R_m$=1, the plasma and the magnetic
field will interact with each other only to a small degree.  
Any weak magnetic disturbances deep in these atmospheres
are unlikely to affect the
surface thermal flux since the winds and weather patterns alluded
to earlier will redistribute the upwelling thermal flux before
it is radiated.   Thus, throughout the atmospheres of L dwarfs,
spanning the range from roughly L2 to L8 \citep{kir99,kir00} we expect
little or no interaction between the visible atmosphere and the magnetic field.

As mentioned above, the choice of $l$ and $v$ are not obvious.  It is easy
to imagine the use of other values, both larger and smaller, for these 
parameters.  Larger values would increase $R_m$ and smaller values would, of 
course, decrease it.  Nonetheless, the values used here are reasonable and 
meant only to estimate $R_m$.

It is also important to note that we have made no assumptions regarding how
magnetic fields in L dwarfs are made or sustained.  Despite our comparisons
to the sun, there is no reason to expect a magnetic field in an L dwarf to be 
created in the same method as the field in the sun.  Indeed, the origin of 
the magnetic field is irrelevant to the calculation above, since it is the
characteristics of the gas that regulate the degree of coupling.

\subsection{Clouds}
The thermal fluxes emerging from L-dwarf atmospheres are affected by clouds.
Early L dwarfs have relatively thin clouds high in the
atmosphere \citep{ack01}.  Figure~\ref{fig-rey} illustrates the trend for 
the clouds to form progressively
deeper in the atmosphere at later spectral types.  The
base of the photosphere is marked for the models shown.
Unlike the case for $R_m$ approaching unity, the clouds form in the immediate
vicinity of the photosphere and thus are well placed to affect the emitted 
thermal flux. For objects with
$T_{\rm eff} \le2000 K$ \citep[approximately L2 and cooler;][]{ste01}
clouds play an important role in the emitted flux since the 
condensate opacity is significant.

For an arbitrary L dwarf the emitted flux in some spectral regions
will be limited by the cloud deck, while in others gaseous opacity
reaches optical depth unity above the cloud.  \citet{marl02}
illustrate the effects of gaseous and condensate opacity for
a variety of L dwarfs.  If there is a transitory clearing in the cloud
deck additional flux will emerge from those spectral regions in which
the cloud opacity is otherwise dominant.  Examples are the peaks of flux
emerging from the water band windows in {\it z, J, H,} and $K$ bands and
the optical flux in $I$ band.  The resulting bright spots on the
objects will be similar to the `5-$\mu$m hot spots' of Jupiter
\citep{wes74} where flux emerges from holes in the ammonia cloud.
An atmosphere with such non-uniformly distributed high contrast regions
should be quite capable of producing photometric variations.

As cloud optical thickness increases with later spectral type, 
L-dwarf $J-K_s$ color becomes redder,
eventually saturating around 2 \citep{marl00,all01,ack01,marl02,tsu01}.
Clouds in cooler objects lie well below the photosphere, and 
leave the radiating region in the atmosphere 
relatively clear.  The clear atmosphere partially manifests itself
in the blue $J-K_s$ seen in the T dwarfs \citep{all96,marl96,tsu96}.

Models predict \citep{marl00,marl02,burg02}
that a hypothetical L dwarf with no clouds will be substantially bluer
\citep[$\sim1.5$ mag;][]{marl01}
at $J-K_s$ than a more realistic object with the same effective
temperature and a cloudy atmosphere.  Thus, if the average L dwarf at
a given spectral type is entirely covered with clouds, it would
not likely be seen as a variable and it would have more typical 
$J-K_s$ color.  In order for photometric variations to arise by
the cloud mechanism there must be
non-uniformity in the cloud coverage, such as clearings in the clouds.
So, not only would clear sections of the atmosphere (holes) provide a source
for brightness variations, flux emerging through such
holes would cause the $J-K_s$ color to be somewhat bluer
than the average object.  The lack of a similar trend for the
early L and late M dwarfs might indicate a different mechanism is
at work in those atmospheres.  Of course variability caused by thicker clouds
would result in variable objects being redder in $J-K_s$ than non-variable
objects.  

\begin{figure}
\plotone{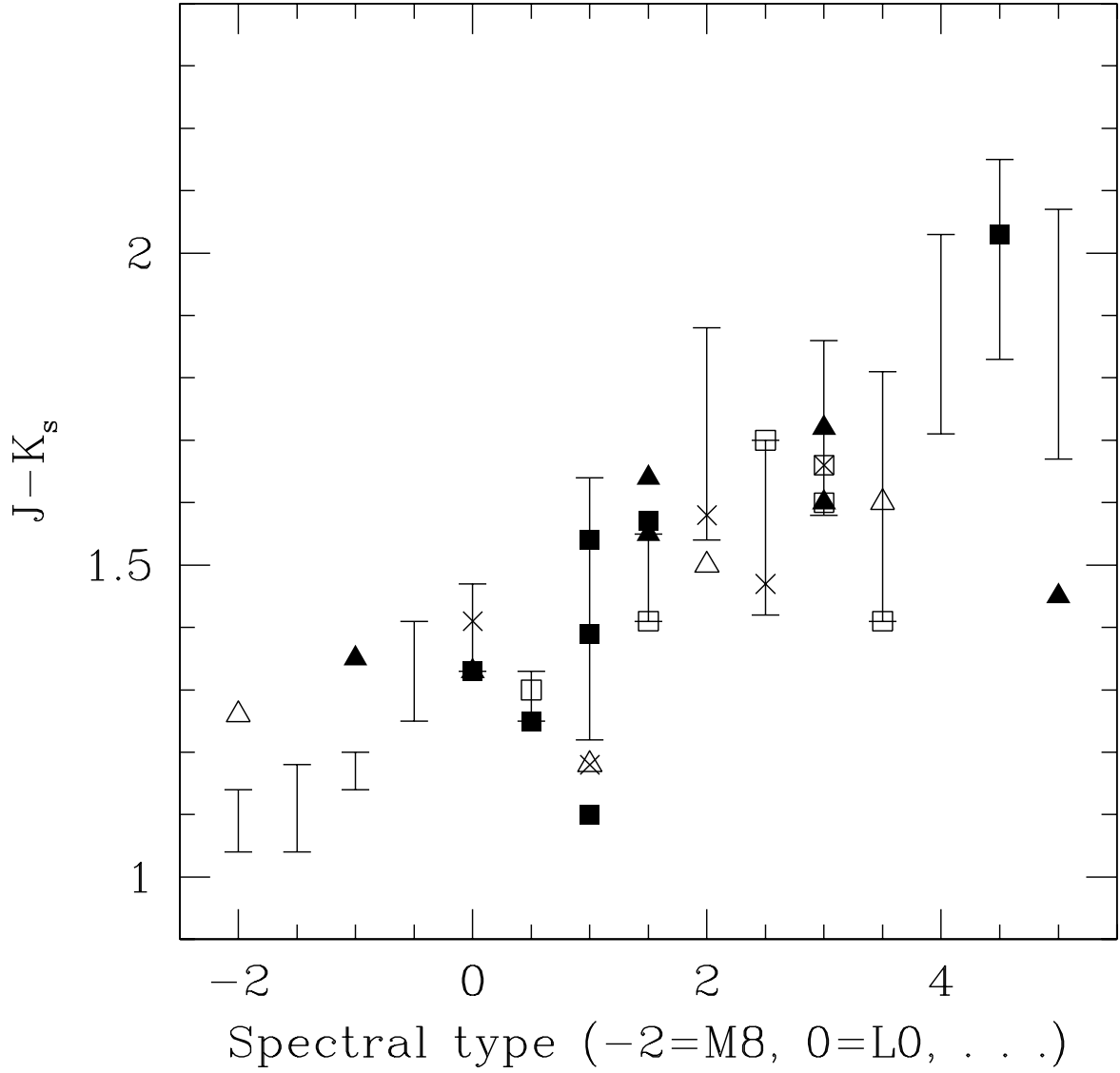}
\caption{2MASS $J-K_s$ color as a function of spectral type.  The symbols are
the same as in Figure~\ref{hafig}.  No clear trends are seen that distinguish
the variable objects from the non-variable objects.
  \label{jkfig}}
\end{figure}

Time-resolved multicolor photometry should be sufficient
to determine if variability is indeed connected to color.  In the absence of
such observations, we examine the previously published $J-K_s$ colors in 
search of systematic differences between variable and non-variable L dwarfs.
Figure~\ref{jkfig} shows that there is no distinct difference between the
variable and non-variable targets.  Incidentally, plots using $J-H$ and
$H-K_s$ also fail to reveal any significant trends \citep{gel02}.  
Perhaps this indicates 
that the cloud features producing the variations are not as simple as a 
single, isolated spot.

\citet{sch00} argue that the atmospheres of L dwarfs likely exist in one of two
states, chaotic or banded.  Since clouds respond to atmospheric motions,
they would presumably reflect one of these two morphologies.
In general, Schubert \& Zhang expect higher mass objects to have more
chaotic and three-dimensional internal dynamics than lower mass objects,
meaning that the higher mass objects are less likely to have banded
cloud features.  It is not obvious which cloud morphology would better
produce photometric variations.  For this reason, it is useful to speculate
on what effects different cloud morphologies can have on an object's 
photometry.  

Objects with more chaotic atmospheres, for example, might be more likely to 
have uniformly distributed
clouds. When rotating, such objects might show little photometric variation,
unless chaotic motion produces rapid evolution of the clouds.  
If the chaotic atmospheres result in fairly
complete cloud coverage, then we might expect that more massive L
dwarfs would be redder in $J-K_s$ and tend not to be variable.  
Conversely, if chaotic atmospheres more often produce large
clearings in the clouds, then they may more easily produce photometric
signatures than banded atmospheres.  In this scenario, the more
massive L dwarfs would be bluer in $J-K_s$ and tend to be variable.
Furthermore, rapidly evolving, chaotic
atmospheres may be responsible for the changes in photometric
period observed for some objects.  It is easy to imagine similar
scenarios for banded clouds.  The lack of any trends with color
for the few L dwarfs observed for variability could indicate that a variety
of cloud morphologies are present in these objects.  Clearly, more 
observations and modeling are required to better characterize atmospheric 
circulation and weather in L-dwarf atmospheres.

\section{Discussion}
The presence of clouds can explain the different types of variability
seen in L dwarfs: non-variable, non-periodic variable, and periodic variable.
The non-variable L dwarfs could be those objects whose atmospheres are either
completely covered with clouds or whose clouds are uniformly distributed
in spots and bands.  
Such cloud morphologies result in small photometric variations below our 
detection limit.

For an object to be classified as a periodic variable it must have some large
feature that produces photometric variations and is stable, both temporally
and spatially, over the entire
span of the observations.  None of the variable L dwarfs exhibit any 
periodic variations that lasted the entire program, suggesting that these 
conditions are not satisfied 
simultaneously in these objects.  For example, a large storm could migrate 
latitudinally or dissipate and reform at a different latitude
as does the Great Dark Spot in the atmosphere of Neptune \citep{ham97}.  
If wind speed is a function of latitude as on all the giant
planets of our solar system, then spots
at different latitudes will circle the object with different periods.

If the migrations or dissipations occur on timescales much shorter than the 
baselines of our observations, then we would not be able to find any 
periodic signal.  On the
other hand, if the evolutionary timescales are a significant fraction of the
baseline, then a prominent feature should produce a temporary signal, though
not necessarily periodic.  Only two objects possibly fall under the latter 
category, 2MASS 0746+20AB and 2MASS 1300+19.  They show features changing on 
timescales about
10 days and longer.  The shortest baseline for the other variable objects is 
around 25 
days.  Any evolution of clouds in their atmospheres must occur on timescales 
shorter than this.  Indeed, \citet{bai01a} conclude that features they observe
must evolve on timescales 
shorter than their maximum observing duration of around 5 days.  

If an L dwarf has several large features at different latitudes, the 
photometric variations produced could be non-periodic.
As mentioned above, an L dwarf is possibly a differentially rotating object.  
Features located at different latitudes may respond to different wind speeds,
resulting in different 
rotation periods around the object.  The resulting light curve would be a
complex composite of several periodic signals with different amplitudes,
periods, and phases.  Photometric ``noise'' from smaller cloud features would
degrade the periodic signals further, essentially making them undetectable.
Since most of the light curves for the objects studied here have data 
randomly scattered about a mean value, it is likely that
the atmospheres for these objects are non-uniformly covered with clouds 
and do not have any single dominant feature.

\section{Conclusions}
We have conducted a photometric monitoring program of eighteen L dwarfs in the
$I_C$ photometric filter.  
We find that seven of these eighteen display statistically significant
variations above the 95.4\% probability level and with RMS scatter between
0.010 and 0.083 mag.  An additional five objects have probabilities for 
variability $\approx$80\%, suggesting that these objects are possibly
variable.  The remaining targets have probabilities $<$35\%, indicating
that they are non-variable or have amplitudes below our detectability.

Only three variable L dwarfs have prominent periods in their CLEAN
power spectra that have power $>$3 times the noise.  The 24.1-hour
period of 2MASS 0345+25 is not intrinsic to the L dwarf and needs more 
investigation.  2MASS 1300+19 has a 9-day section of its light curve that 
could be evidence for the creation and dissipation of a large storm.  The 
source of the 31.0-hour period of 2MASS 0746+20AB 
is not clear from the light curve.  However, it does have a 70-day section with
a puzzling saw-tooth pattern with properties unlike any that would be expected.
All of these timescales are much longer than the rotation periods expected 
for these objects.

The light curves of the other variable objects are quite random and have no 
dominant features.  Non-uniform cloud coverage with features evolving on
timescales less than a few days is likely the source of the variations in these
objects.  Indeed, we have shown that the low
ionization fraction predicted by L-dwarf models and the accompanyingly
low magnetic Reynolds numbers strongly argue against
magnetic spots as a plausible cause for the photometric variations.  
On the other hand silicate and iron grains condense in L-dwarf atmospheres 
within the photosphere.  These clouds are likely responsible for the 
photometric variations discovered in the various studies, particularly for
the later L dwarfs (about L2 and later).  Since the thermal emission
of T dwarfs is also influenced by clouds \citep{marl02} we predict
that variability will also be found in the opacity window regions of these
objects. Further work with models and more observations are required to better
understand cloud composition and dynamics.

\begin{acknowledgements}
The authors wish to thank D.H. Roberts, J. Peterson, and B. Goldman for 
various contributions and discussions.  We also wish to acknowledge the
useful comments of the referee.
C.G. and M.M. acknowledge support from NASA grants NAG2-6007 and NAG5-8919
and  NSF grants AST-9624878 and AST-0086288. 
Work by K.L. supported by NSF grant AST-0086487.
\end{acknowledgements}

\begin{deluxetable}{lcccccc}
\tablecolumns{7}
\tablewidth{0pc}
\tablecaption{L Dwarfs Observed for Variability \label{tab-list}}
\tablehead{
  \colhead{Name}
 &\colhead{Spectral}
 &\colhead{R.A.}
 &\colhead{Dec.}
 &\colhead{$I$}
 &\colhead{H$\alpha$}
 &\colhead{Reference} \\
  \colhead{}
 &\colhead{Type}
 &\colhead{J(2000)}
 &\colhead{J(2000)}
 &\colhead{Magnitude\tablenotemark{a}}
 &\colhead{Emission\tablenotemark{b}}
 &\colhead{}
}

\startdata
2MASS 0015+35 & L2   & 00:15:44.7 & +35:16:03 & $\sim17.2$ & Y & 1 \\
2MASS 0036+18 & L3.5 & 00:36:15.9 & +18:21:10 & 16.10      & \nodata & 1,2 \\
2MASS 0058$-$06 & L0   & 00:58:42.5 & $-$06:51:23 & $\sim17.5$ & Y & 1 \\
2MASS 0135+12 & L1.5 & 01:35:35.8 & +12:05:22 & $\sim17.7$ & Y & 1 \\
2MASS 0345+25 & L0   & 03:45:43.2 & +25:40:23 & 16.98      & \nodata & 3 \\
2MASS 0746+20AB & L0.5 & 07:46:42.5 & +20:00:32 & 15.11      & Y & 1,2 \\
2MASS 1029+16 & L2.5 & 10:29:21.6 & +16:26:52 & $\sim17.6$ & Y & 1 \\
2MASS 1108+68 & L1   & 11:08:30.7 & +68:30:17 & $\sim16.6$ & Y & 4 \\
2MASS 1146+22AB & L3   & 11:46:34.5 & +22:30:53 & 17.62      & \nodata & 3 \\
2MASS 1300+19 & L1   & 13:00:42.5 & +19:12:35 & $\sim15.9$ & \nodata & 4 \\
2MASS 1338+41 & L2.5 & 13:38:26.1 & +41:40:34 & $\sim17.6$ & \nodata & 1 \\
2MASS 1411+39 & L1.5 & 14:11:17.5 & +39:36:36 & $\sim17.9$ & \nodata & 1 \\
2MASS 1412+16 & L0.5 & 14:12:24.4 & +16:33:12 & $\sim17.1$ & Y & 1 \\
2MASS 1439+19 & L1   & 14:39:28.4 & +19:29:15 & 16.02      & \nodata & 2,3 \\
2MASS 1506+13 & L3   & 15:06:54.4 & +13:21:06 & $\sim16.9$ & Y & 4 \\
2MASS 1615+35 & L3   & 16:15:44.1 & +35:59:00 & $\sim18.1$ & \nodata & 1 \\
2MASS 1658+70 & L1   & 16:58:03.7 & +70:27:01 & $\sim16.7$ & \nodata & 4 \\
2MASS 2224$-$01 & L4.5 & 22:24:43.8 & $-$01:58:52 & $\sim18.0$ & Y & 1
\enddata
\tablenotetext{a}{Approximate magnitudes are estimated from the $I-J$ color
and the instrumental magnitudes.}
\tablenotetext{b}{All targets have been observed for H$\alpha$ emission.  
Those entries listed as `Y' have had definite detections;
entries listed as `\nodata' indicate that only upper limits for emission
have been obtained.}
\tablerefs{(1) \citealt{kir00}; (2) \citealt{reid00}; (3) \citealt{kir99};
(4) \citealt{giz00}}
\end{deluxetable}

\begin{deluxetable}{lccccccc}
\tablecolumns{8}
\tablewidth{0pc}
\tablecaption{L-Dwarf Variability Results \label{tab-sv}}
\tablehead{
  \colhead{Name}
 &\colhead{Nights}
 &\colhead{Frames}
 &\colhead{t$_{max}$}
 &\colhead{References}
 &\colhead{$p$}
 &\colhead{$\sigma_{\rm RMS}$}
 &\colhead{$\overline{\sigma_m}$}  \\
  \colhead{}
 &\colhead{Used/Total}
 &\colhead{Used/Total}
 &\colhead{[days]}
 &\colhead{Used/Total}
 &\colhead{[\%]}
 &\colhead{[mag]}
 &\colhead{[mag]}
}

\startdata
2MASS 0345+25\tablenotemark{a} & 35/37 & 209/318 & 108.73 &7/7& $>$99.99& 0.030 & 0.027\\
2MASS 0746+20AB&38/45 & 260/421 & 150.88 & 10/12& $>$99.99 & 0.010 & 0.007 \\
2MASS 1300+19 & 37/39 & 275/384 & 137.78 &  5/7 & $>$99.99 & 0.015 & 0.012 \\
2MASS 2224$-$01 &  5/6  &  15/21  & 26.01  &  5/7 & 99.88 & 0.083 & 0.057 \\
2MASS 1108+68 & 46/52 & 394/536 & 182.80 &  5/8 & 99.75 & 0.016 & 0.016 \\
2MASS 0135+12 & 18/22 & 110/187 & 30.00  &  5/6 & 99.33 & 0.041 & 0.035 \\
2MASS 1658+70 & 10/13 &  27/38  & 31.99  & 9/14 & 97.31 & 0.024 & 0.019 \\
\hline
2MASS 1615+35 & 12/15 &  43/61  & 40.92  &  6/6 & 82.69 & 0.073 & 0.067 \\
2MASS 0015+35 &  5/9  &  21/34  & 230.31 &  6/6 & 80.82 & 0.026 & 0.024 \\
2MASS 0058$-$06 &  3/3  &  11/12  & 26.95  &  5/5 & 80.23 & 0.034 & 0.028 \\
2MASS 1439+19\tablenotemark{b} & 11/11 &  28/30  & 38.90  & 4/4 & 79.81 & 0.014 & 0.013 \\
2MASS 1338+41 & 27/29 & 112/140 & 123.86 &  8/9 & 77.10 & 0.039 & 0.035 \\
\hline
2MASS 1029+16 &   5/5 &  14/18  & 32.96  & 6/6 & 30.39 & 0.057 & 0.063 \\
2MASS 0036+18 & 10/18  & 36/85  & 53.00  & 6/8 & 26.66 & 0.009 & 0.010 \\
2MASS 1506+13 &   3/4 &   9/12  & 35.91  & 5/6 & 22.46 & 0.024 & 0.030 \\
2MASS 1412+16 &  8/11 &  27/39  & 84.92  & 4/4 & 11.40 & 0.018 & 0.025 \\
2MASS 1146+22AB\tablenotemark{a} &10/16 &  28/48  & 36.90  & 5/5 & 0.54 & 0.034 & 0.057 \\
2MASS 1411+39 & 24/30 & 78/132  & 131.86 & 4/5 & 0.12 & 0.047 & 0.056
\enddata
\tablenotetext{a}{\citet{bai01a} detected variations in these objects in their
study.}
\tablenotetext{b}{\citet{bai01a} did not detect variations in this object in 
their study.}
\end{deluxetable}

\clearpage

\begin{deluxetable}{lcc}
\tablecolumns{4}
\tablewidth{0pc}
\tablecaption{CLEAN Results for Variable and Possibly Variable Targets 
  \label{tab-pow}}
\tablehead{
   \colhead{Object}
  &\colhead{Primary Peak}
  &\colhead{PNR\tablenotemark{a}} \\
   \colhead{}
  &\colhead{Period [hours]}
  &\colhead{}
}

\startdata
2MASS 0746+20AB & 31.0$\pm$0.1  & 4.85 \\
2MASS 1300+19   & 238.$\pm$9.   & 3.38 \\
2MASS 1108+68   & 23.8$\pm$0.1  & 2.55 \\
2MASS 1658+70   & 1.89$\pm$0.01 & 1.64 \\
2MASS 0135+12   & 18.6$\pm$0.2  & 1.48 \\
2MASS 2224$-$01 & 21.8$\pm$0.4  & 0.71 \\
\hline
2MASS 1439+19   & 2.60$\pm$0.01 & 1.85 \\
2MASS 1615+35   & 1.01$\pm$0.01 & 1.24 \\
2MASS 1338+41   & 6.68$\pm$0.01 & 1.06 \\
2MASS 0015+35   & 9.12$\pm$0.01 & 0.85 \\
2MASS 0058$-$06 & 2.34$\pm$0.10 & 0.74 
\enddata
\tablenotetext{a}{PNR = ratio of CLEAN peak power to the noise.}
\end{deluxetable}

\clearpage

\clearpage

\clearpage

\end{document}